\documentclass[a4paper,11pt]{article}
\pdfoutput=1 

\usepackage{jheppub} 
\usepackage[T1]{fontenc} 

\usepackage{graphicx}  
\usepackage{dcolumn}   
\usepackage{bm}        
\usepackage{amssymb}   
\usepackage{amsmath}

\usepackage{hyperref}

\usepackage{hhline}
\usepackage{xcolor}

\newcommand{\D}{\mathrm{d}} 
\renewcommand\vec[1]{\ensuremath\boldsymbol{#1}} 
\DeclareMathOperator{\Tr}{Tr}  

\title{\boldmath A title with some math: $x=1$}

\title{Emergent chiral symmetry in a three-dimensional interacting Dirac liquid}

\author[a]{Andr\' as L. Szab\' o,}
\author[b]{Bitan Roy}

\affiliation[a]{Max-Planck-Institut f\"{u}r Physik komplexer Systeme, N\"{o}thnitzer Str. 38, 01187 Dresden, Germany}
\affiliation[b]{Department of Physics, Lehigh University, Bethlehem, Pennsylvania, 18015, USA}

\date{\today}

\emailAdd{szabo@pks.mpg.de}
\emailAdd{bitan.roy@lehigh.edu}

\abstract{We compute the effects of strong Hubbardlike local electronic interactions on three-dimensional four-component massless Dirac fermions, which in a noninteracting system possess a microscopic global U(1)$\otimes$SU(2) chiral symmetry. A concrete lattice realization of such chiral Dirac excitations is presented, and the role of electron-electron interactions is studied by performing a field theoretic renormalization group (RG) analysis, controlled by a \emph{small} parameter $\epsilon$ with $\epsilon=d-1$, about the lower-critical one spatial dimension. Besides the noninteracting Gaussian fixed point, the system supports four quantum critical and four bicritical points at nonvanishing interaction couplings $\sim \epsilon$. Even though the chiral symmetry is absent in the interacting model, it gets restored (either partially or fully) at various RG fixed points as emergent phenomena. A representative cut of the global phase diagram displays a confluence of scalar and pseudoscalar excitonic and superconducting (such as the $s$-wave and $p$-wave) mass ordered phases, manifesting restoration of (a) chiral U(1) symmetry between two excitonic masses for repulsive interactions and (b) pseudospin SU(2) symmetry between scalar or pseudoscalar excitonic and superconducting masses for attractive interactions. Finally, we perturbatively study the effects of \emph{weak} rotational symmetry breaking on the stability of various RG fixed points.}

\keywords{Dirac fermions, Chiral symmetry, Renormalization group, $\epsilon$ expansion.}

\arxivnumber{2009.05055}

\begin{document}

\maketitle
\flushbottom

\section{Introduction}

Dirac fermions offer a universal language to explore various territories of modern physics that include baryonic matters at high energies~\cite{peskin-schroeder}, topological phases of matter~\cite{Shen-book, Bernevig-book}, and scale invariant quantum critical phenomena in fermionic systems, to name a few. While in the context of high-energy physics \emph{massless} Dirac fermions are realized in the ultraviolet regime, in solid state systems they are found as emergent sharp quasiparticle excitations at sufficiently low energies (the infrared regime)~\cite{balatsky:review, armitage:RMP}. The prominent representatives of such infrared nodal Dirac materials are the two-dimensional carbon-based honeycomb membrane or graphene~\cite{graphene:RMP}, and three-dimensional strong spin-orbit coupled Cd$_3$As$_2$~\cite{cd2as3:Exp} and Na$_3$Bi~\cite{na3bi:Exp}. Here we focus on the minimal building block of a three-dimensional strong spin-orbit coupled gapless Dirac liquid, constituted by four-component massless Dirac fermions, that besides the fundamental discrete parity (${\mathcal P}$), time-reversal (${\mathcal T}$) and charge-conjugation (${\mathcal C}$) symmetries, also enjoy a \emph{microscopic} continuous U(1)$\otimes$SU(2) global chiral symmetry. The present discussion unfolds the role of strong momentum-independent Hubbardlike or local electronic interactions among massless chiral Dirac fermions.

The central outcomes of this study are captured by a representative cut of the global phase diagram, shown in Fig.~\ref{fig:phasediag}. It displays realizations of and competitions among \emph{four} chiral symmetry breaking \emph{mass} orders for sufficiently strong local electronic interactions, which we demonstrate via a controlled, but leading-order renormalization group (RG) analysis. Even though the interacting Hubbardlike model does not possess the chiral U(1)$\otimes$SU(2) symmetry at the bare level, the resulting phase diagram and the RG fixed points manifest its restoration (at least partially) as emergent phenomena. Indeed ``\emph{more is different}"~\cite{anderson}. We now present a synopsis of our main findings.

\subsection{Summary of results}

Here we show that a collection of isotropically dispersing massless chiral Dirac fermions, interacting via Hubbardlike local or short-range interactions, is described only in terms of \emph{four} linearly independent four-fermion or quartic terms. By performing a one-loop or leading-order RG analysis on such an interacting model, controlled by a \emph{small} parameter $\epsilon$ with $\epsilon=d-1$, about the lower-critical one spatial dimension ($d=1$), we show that the system altogether sustains eight interacting fixed points, see Table~\ref{tab:FPs}. All the fixed points are located at coupling constants $\sim \epsilon$, and can be grouped into following three categories, depending on the emergent chiral symmetry therein. Fixed points (1) possessing a partial U(1) or pseudospin SU(2) chiral symmetry, (2) enjoying the full U(1)$\otimes$ SU(2) chiral symmetry, and (3) transforming into each other under the chiral U(1) rotation. We arrive at these conclusions by computing the scaling dimensions of all symmetry allowed particle-hole or excitonic and particle-particle or superconducting orders, see Table~\ref{tab:bilinears}, at various RG fixed points, see Table~\ref{tab:AnomDim}. Otherwise, among eight RG fixed points four are quantum critical points (QCPs), while the remaining ones are bircitical points (BCPs). All the QCPs are characterized by the dynamic scaling exponent $z=1$ and the correlation length exponent $\nu=\epsilon^{-1}$. Therefore, at the upper-critical three spatial dimensions we recover the \emph{exact} mean-field value of the exponent $\nu=1/2$~\cite{zinn-justin:book, zinn-justin-moshe-moshe}, from a leading order $\epsilon$ expansion.

\begin{figure}[tbp]
\centering
\includegraphics[width=0.5\linewidth]{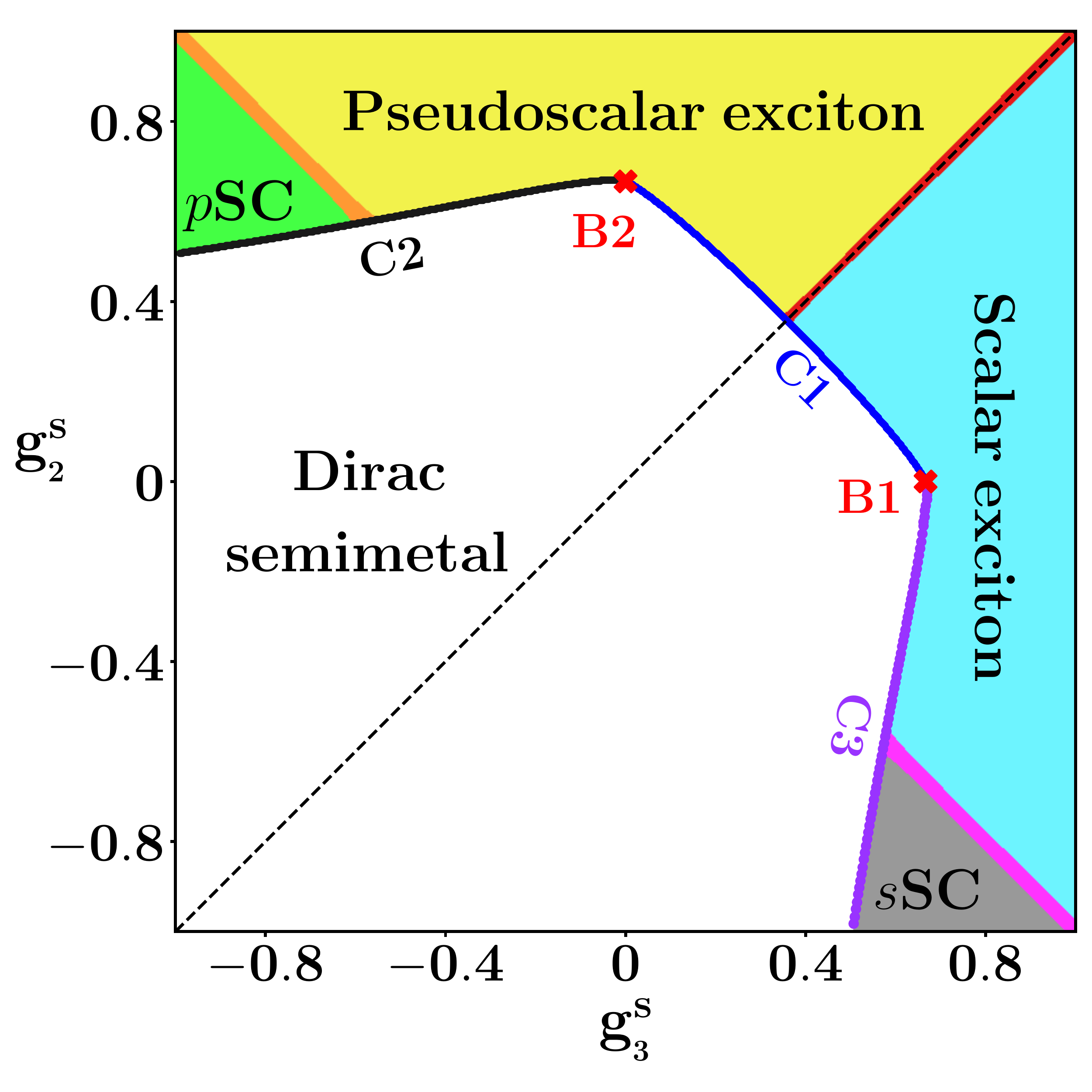}
\caption{A representative phase diagram of a three-dimensional interacting isotropic Dirac semimetal in a two-dimensional subspace of interaction couplings. Here $g^s_{_3}$ and $g^s_{_2}$ are the coupling constants in scalar and pseudoscalar excitonic mass channels, respectively, measured in units of $\epsilon=d-1$. Their positive (negative) values correspond to repulsive (attractive) interactions. Together, these two masses form a U(1) vector under the chiral U(1) rotation. The scalar $s$-wave and pseudoscalar $p$-wave superconducting masses (denoted by $s$SC and $p$SC, respectively) also constitute a U(1) chiral vector. The emergent chiral symmetry manifests through the \emph{mirror symmetry} of the phase diagram about the $g^s_{_3}=g^s_{_2}$ line (the diagonal dashed line). Under the mirror transformation the components of the chiral U(1) mass orders transform into each other. The Dirac semimetal-excitonic mass quantum phase transition (blue boundary) is governed by the U(1) symmetric quantum critical point (QCP) C1. The scalar and pseudoscalar excitonic masses are degenerate along the red phase boundary between them, where the ordered state represents a ${\mathcal P}$ and ${\mathcal T}$ symmetry breaking \emph{axionic} insulator~\cite{peccei-quinn, weinberg, wilczek}. The phase transitions across the black and purple boundaries are governed by the pseudospin SU(2) symmetric QCPs C2 and C3, respectively. Each excitonic mass is degenerate with the adjacent superconducting mass along the 135$^\circ$ diagonal (orange and pink lines), where they constitute pseudospin SU(2) chiral vectors, see Fig.~\ref{fig:pseudospin}. The basins of attraction of C1 and C2 (C1 and C3) are separated by the bicritical point B2 (B1), see Table~\ref{tab:FPs}. The transitions through the bicritical points are also continuous, as they are accessed by holding one of the unstable directions fixed~\cite{roy-foster:PRX}.  
}~\label{fig:phasediag}
\end{figure}

Some of these fixed points also play prominent roles on the global phase diagram of interacting Dirac fermions, a representative cut of which is displayed in Fig.~\ref{fig:phasediag}. It shows a competition and confluence of \emph{four} mass orders for chiral Dirac fermions, namely (1) the scalar excitonic mass, (2) the pseudoscalar excitonic mass, (3) the scalar $s$-wave pairing and (4) the pseudoscalar odd-parity $p$-wave pairing. Transformations of each mass order under various discrete (${\mathcal P}$, ${\mathcal T}$ and ${\mathcal C}$) and continuous symmetries are summarized in Table~\ref{tab:bilinears}. The fixed points (both critical and bicritical) controlling the continuous quantum phase transitions out of a Dirac semimetal into various broken symmetry phases for finite interactions across different segments of the phase boundary are also highlighted in Fig.~\ref{fig:phasediag}. Most importantly, we find that the U(1) chiral symmetry of the noninteracting system manifests in the phase diagram as a reflection about the 45$^\circ$ diagonal (dashed) line under which the scalar excitonic and pairing masses transform into the pseudoscalar masses. Along this high symmetry line the chiral U(1) symmetry between scalar and pseudosclar excitonic masses gets restored in the ordered phase (red line), which describes an \emph{axionic} insulator~\cite{peccei-quinn, weinberg, wilczek}. Furthermore, the phase diagram supports another high symmetry line, the 135$^\circ$ diagonal one, along which the pseudospin SU(2) symmetry between the scalar (pink line) and pseudoscalar (orange line) excitonic and pairing masses gets restored.

%
\begin{table}[t!]
\centering
\begin{tabular}{|c|c|c|c|c|c|c|c|c|}
\hline
coupling & C1 & C2 & C3 & C4 & B1 & B2 & B3 & B4 \\
\hline \hline
$g^s_{_0}$ & -0.042 & 0.062  &  0.062 & -2/3 &  0  &  0  & -0.895 & -0.895 \\
$g^s_{_1}$ & -0.125 & 0.136  &  0.136 &   0  &  0  &  0  &  0.614 &  0.614 \\
$g^s_{_2}$ &  0.167 & 0.215  & -0.153 & -1/3 &  0  & 1/3 & -0.056 & -0.840 \\
$g^s_{_3}$ &  0.167 & -0.153 &  0.215 & -1/3 & 1/3 &  0  & -0.840 & -0.056 \\
\hline
\end{tabular}
\caption{Locations of four quantum critical points (QCPs) [C1, $\cdots$, C4] and four bicritical points (BCPs) [B1, $\cdots$, B4] in the four-dimensional space of coupling constants, measured in units of $\epsilon$, where $\epsilon=d-1$ and $d$ is the spatial dimension of the system. Each QCP (BCP) possesses \emph{one} (\emph{two}) unstable direction(s). Note C1 and C4 are chiral U(1) symmetric QCPs, since $g^s_{_2}=g^s_{_3}$ therein. On the other hand, C2 and C3 are chiral U(1) partners, as their locations transform into each other under U(1) chiral rotation $g^s_{_2} \leftrightarrow g^s_{_3}$. Similarly, two pairs of BCPs, namely (1) (B1,B2), and (2) (B3,B4) are chiral U(1) partners. In addition, three QCPs C2, C3 and C4, and two BCPs B3 and B4 individually possess pseudospin SU(2) chiral symmetry. The emergent chiral symmetries can be anchored by comparing the scaling dimensions of fermion bilinears, tabulated in Table~\ref{tab:bilinears}, at various fixed points, see Table~\ref{tab:AnomDim}. In particular, the scaling dimensions of fermion bilinears that are related to each other by chiral U(1) [pseudospin SU(2) chiral] rotations are identical at the U(1) [pseudospin SU(2)] symmetric fixed points. By contrast, the scaling dimensions of two fermion bilinears that form a U(1) vector under the chiral U(1) rotations, are interchanged between two fixed points that are chiral partners of each other. Finally, the scaling dimension of chiral U(1) scalar fermion bilinears remain unchanged between two chiral partner fixed points. Only the fully U(1)$\otimes$SU(2) symmetric QCP C4 becomes unstable even against weak rotational symmetry breaking, see Sec.~\ref{subsec:anisotropy}.        
}~\label{tab:FPs}
\end{table}

Even though here we arrive at these conclusions from a leading-order RG analysis within the framework of an $\epsilon$ expansion about the lower-critical one spatial dimension, they are expected to be valid in general for the following reasons. Notice that existence of the fixed points do not depend on the value of $\epsilon$, as long as $\epsilon=d-1>0$. Moreover, their locations (in terms of the coupling constants) are quoted in units of $\epsilon$ (Table~\ref{tab:FPs}). Consequently, the scaling dimension for all the fermion bilinears (Table~\ref{tab:bilinears}) are also reported in units of $\epsilon$ (Table~\ref{tab:AnomDim}). Finally, the chiral symmetry among various competing phases, which we demonstrate as an emergent phenomena from a leading-order $\epsilon$ expansion, is, however, an exact symmetry, as shown in Fig.~\ref{fig:pseudospin}.

Finally, we address the stability of various interacting fixed points against \emph{weak} breakdown of the spatial rotational symmetry. Such a rotational symmetry breaking introduces an anisotropy between the Fermi velocities in the $xy$ plane and along the $z$ direction, for example. Here we show that all but one RG fixed points, reported in Table~\ref{tab:FPs} for the isotropic system, retain their character in a weakly anisotropic Dirac semimetal. Only one QCP, namely C4, becomes a BCP, see Sec.~\ref{subsec:anisotropy}.

\subsection{Organization}

The rest of the paper is organized as follows. In the next section we present the low-energy model for four-component massless chiral Dirac fermions, discuss its symmetry properties and a lattice realization of such gapless excitations on a cubic lattice. Sec.~\ref{sec:eeinteraction} is devoted to address the effects of strong local electronic interactions, emergent chiral symmetry and quantum critical behavior in this system. In Sec.~\ref{sec:summary} we summarize our findings and allude to some possible future directions. Additional technical details are relegated to the Appendices.

\begin{table}[t!]
\centering
\begin{tabular}{|c|c|c|c|c|c|c|c|}
\hline
CF & Matrix & Physical meaning & ${\mathcal P}$ & ${\mathcal T}$ & ${\mathcal C}$ & CH & PS \\
\hline \hline
$\Delta_0^s$ & $\eta_3\Gamma_{00}$ & Fermionic density & $+$ & $+$ & $-$ & & III \\
$\Delta_1^s$ & $\eta_3\Gamma_{10}$ & Chiral density & $-$ & $+$ & $+$ & & \\
$\Delta_2^s$ & $\eta_0\Gamma_{20}$ & Pseudoscalar mass & $-$ & $-$ & $+$ & \textcolor{red}{$\bullet$} & II \\
$\Delta_3^s$ & $\eta_3\Gamma_{30}$ & Scalar mass & $+$ & $+$ & $+$ & \textcolor{red}{$\bullet$} & I \\
\hline \hline
$\Delta_0^t$ & $\eta_0\Gamma_{0j}$ &  Axial current & $+$ & $-$ & $+$ & & \\
$\Delta_1^t$ & $\eta_0\Gamma_{1j}$ &  Abelian current & $-$ & $-$ & $-$ & & IV \\
$\Delta_2^t$ & $\eta_3\Gamma_{2j}$ &  Spatio-temporal tensor & $-$ & $+$ & $-$ & \textcolor{black}{$\bullet$} & \\
$\Delta_3^t$ & $\eta_0\Gamma_{3j}$ & Spatial tensor & $+$ & $-$ & $-$ & \textcolor{black}{$\bullet$} & \\
\hline \hline
$\Delta_0^p$ & $\eta_\alpha\Gamma_{00}$ & Scalar $s$-wave pairing & $+$ & $+/-$ & $+$ & \textcolor{blue}{$\bullet$} & I \\
$\Delta_1^p$ & $\eta_\alpha\Gamma_{10}$ & Pseudoscalar $p$-wave pairing & $-$ & $+/-$ & $+$ & \textcolor{blue}{$\bullet$} & II \\
$\Delta_2^p$ & $\eta_\alpha\Gamma_{2j}$ & Spatial vector pairing & $-$ & $+/-$ & $+$ & & IV \\
$\Delta_3^p$ & $\eta_\alpha\Gamma_{30}$ & Temporal vector pairing & $+$ & $+/-$ & $+$ & & III \\
\hline
\end{tabular}
\caption{Various local (momentum independent) orderings with their conjugate fields (CFs) (first column), the corresponding matrix $\eta_\mu \Gamma_{\nu \rho}$ (second column) associated with the fermion bilinears $\Psi^\dag_{\rm Nam} \eta_\mu \Gamma_{\nu \rho} \Psi_{\rm Nam}$ in the Nambu basis $\Psi_{\rm Nam}$, defined in Eq.~(\ref{eq:Nambudefinition}), and the physical meaning of the orderings (third column). First eight (last four) rows correspond to excitonic (superconducting) orders. In the superconducting channels $\alpha=1$ and $2$, reflecting the U(1) gauge redundancy in the choice of the superconducting phase ($\phi$), see Eq.~(\ref{eq:pairingsingleparticle}). Transformation of each fermion bilinear under discrete ${\mathcal P}$, ${\mathcal T}$, and ${\mathcal C}$ symmetries are shown in the fourth, fifth and sixth columns, respectively. Here, $+$ and $-$ respectively correspond to even and odd. Fermion bilinears transforming as components of three chiral U(1) vectors under the U(1) chiral (CH) rotation are identified with distinct colored circles (red, black and blue). Rest of the bilinears are scalars under chiral U(1) rotation. Fermion bilinears transforming as components of four pseudospin (PS) SU(2) vectors are marked as I, II, III and IV in the eighth column, see Fig.~\ref{fig:pseudospin}. The rest of the fermion bilinears transform as scalars under the pseudospin rotations, as they all commute with $\vec{\mathrm{PS}}$, see Eq.~(\ref{eq:pseudospingenerator}).     
}\label{tab:bilinears}
\end{table}

\section{Non-interacting system}

In this section we introduce a noninteracting gapless Dirac system, possessing a genuine \emph{microscopic} continuous chiral symmetry. First we promote the continuum or low-energy description of such system, and discuss its invariance under various discrete and continuous symmetries. Subsequently, we propose a lattice realization of a genuine chiral Dirac semimetal.

\begin{table}[t!]
\centering
\begin{tabular}{|c|c|c|c|c|c|c|c|c|}
\hline
CF & C1 & C2 & C3 & C4 & B1 & B2 & B3 & B4 \\
\hline \hline
$\Delta_0^s$ & 0 & 0 & 0 & 0 & 0 & 0 & 0 & 0 \\
$\Delta_1^s$ & 0 & 0 & 0 & 0 & 0 & 0 & 0 & 0 \\
$\Delta_2^s$ & {\bf 0.875} & {\bf 0.849} & -0.256 & -1 & 1/2 & {\bf 4/3} & 0.754 & -1.597 \\
$\Delta_3^s$ & {\bf 0.875} & -0.256 & {\bf 0.849} & -1 & {\bf 4/3} & 1/2 & -1.597 & 0.754 \\
\hline \hline
$\Delta_0^t$ & -0.167 & -0.260 & -0.260 & 4/3 & -1/3 & -1/3 & 1.176 & 1.176 \\
$\Delta_1^t$ & 0.500 & -0.136 & -0.136 & 0 & 1/3 & 1/3 & -0.614 & -0.614 \\
$\Delta_2^t$ & -0.042 & -0.147 & 0.221 & 1/3 & 1/6 & -1/6 & 0.363 & 1.147 \\
$\Delta_3^t$ & -0.042 & 0.221 & -0.147 & 1/3 & -1/6 & 1/6 & 1.147 & 0.363 \\
\hline \hline
$\Delta_0^p$ & -0.250 & -0.256 & {\bf 0.849} & -1 & 1/2 & -1/2 & -1.597 & 0.754 \\
$\Delta_1^p$ & -0.250 & {\bf 0.849} & -0.256 & -1 & -1/2 & 1/2 & 0.754 & -1.597 \\
$\Delta_2^p$ & -0.250 & -0.136 & -0.136 & 0 & -1/3 & -1/3 & -0.614 & -0.614 \\
$\Delta_3^p$ & 0 & 0 & 0 & 0 & 0 & 0 & 0 & 0 \\
 \hline
\end{tabular}
\caption{Scaling dimensions (in units of $\epsilon$) of various particle-hole (first eight rows) and particle-particle (last four rows) order parameters (see Table~\ref{tab:bilinears}) at the eight nontrivial fixed points (see Table~\ref{tab:FPs}). At the chiral U(1) symmetric fixed point C1 any two orders transforming as chiral U(1) vector (see Table~\ref{tab:bilinears}, seventh column) possess \emph{identical} scaling dimensions. By contrast, their scaling dimensions \emph{switch} between two fixed points that are chiral U(1) partners of each other, namely (1) (C2,C3), (2) (B1,B2), and (3) (B3,B4). At four pseudospin SU(2) chiral symmetric fixed points (C2, C3, B3, B4) all components of each pseudospin SU(2) vector (see Table~\ref{tab:bilinears}, eighth column and Fig.~\ref{fig:pseudospin}) possess identical scaling dimension. At the fully U(1)$\otimes$SU(2) chiral symmetric fixed point C4, all components of chiral U(1) and pseudospin SU(2) vectors acquire equal scaling dimensions. In the phase diagram shown in Fig.~\ref{fig:phasediag}, we highlight the role of these fixed points. At three QCPs (C1, C2, and C3) and two BCPs (B1 and B2) the mass orders possess the largest scaling dimensions (shown in bold). Therefore, C1 controls the transition to scalar and pseudoscalar excitonic mass orders. By contrast, C2 (C3) controls transition to pseudoscalar (scalar) excitonic and superconducting masses. On the other hand, when the role of C1 and C3 (C1 and C2) switches, the continuous transition to the scalar (pseudoscalar) excitonic mass is controlled by the BCP B1 (B2).  
}~\label{tab:AnomDim}
\end{table}

\subsection{Continuum model and symmetries}

The minimal model for three-dimensional massless chiral Dirac fermions is captured by the Hamiltonian
\begin{equation}~\label{eq:H0}
\hat{h}(\vec{k})= \sum_{j=1}^3 v_j\Gamma_{j}k_j, 
\end{equation}
where $k_j$ are the components of momentum and $v_j$ are the Fermi velocities along the principal axes. Here $j=1,2$ and $3$ correspond to $x$, $y$, and $z$, respectively. The mutually anticommuting four-component Hermitian $\Gamma$ matrices satisfy the Clifford algebra $\{\Gamma_i,\Gamma_j\}=2 \delta_{ij}$. The representation of $\Gamma$ matrices depends on microscopic details, even though our results are insensitive to it. Nonetheless, for the sake of concreteness we organize the four-component spinor basis as $\Psi_{\vec{k}}=(c_{+\uparrow},c_{+\downarrow},c_{-\uparrow},c_{-\downarrow})^\top( \vec{k})$, where $c_{p s}(\vec{k})$ are the fermion annihilation operators with parity $p=\pm$, spin projections $s=\uparrow, \downarrow$, and momentum $\vec{k}$. The $\Gamma$ matrices are then given by $\Gamma_j=\Gamma_{1j}$ for $j=1,2,3$, with $\Gamma_{\mu \nu} \equiv \tau_\mu \otimes \sigma_\nu$. Two sets of Pauli matrices $\{ \tau_\mu \}$ and $\{ \sigma_\nu \}$ operate on parity and spin indices, respectively, and $\tau_0$ and $\sigma_0$ are the two-dimensional identity matrices. The model in Eq.~(\ref{eq:H0}) then describes the mixing of different parity states with unit angular momentum difference. To close the anticommuting Clifford algebra, which contains five elements in the space of four-dimensional Hermitian matrices, we define $\Gamma_4=\Gamma_{30}$ and $\Gamma_5=\Gamma_{20}$.

Unless mentioned, throughout the paper we set $v_x=v_y=v_z=v$ (say), which brings Eq.~(\ref{eq:H0}) into a fully rotationally symmetric form $\hat{h}(\vec{k})=v \Gamma_{1j} k_j$, where the summation over the repeated spatial indices is assumed, and $\hat{h}(\vec{k})$ describes an isotropic, doubly Kramers degenerate Dirac cone centered at the $\Gamma=(0,0,0)$ point [see Sec.~\ref{subsec:latticemodel}]. The generators of rotations about the $x$-, $y$- and $z$-axis are respectively $\Gamma_{01}, \Gamma_{02}$, and $\Gamma_{03}$. For example, a rotation by an angle $\theta$ about the $j$th axis is generated by $R_{j}({\theta})=\exp[i \theta \Gamma_{0j}/2]$. Specifically, when $\theta=\pi/2$ and $j=z$, $\vec{k} \to (k_y,-k_x,k_z)$, such that $R_{z}({\pi/2})\hat{h}(\vec{k})R^{-1}_{z}({\pi/2})=\hat{h}(\vec{k})$ in an isotropic system. Therefore, the noninteracting Hamiltonian remains invariant under the four-fold ($C_4$) rotation about the $z$ axis. Similarly, under the $C_4$ rotations about the $x$ and $y$ axes $\vec{k} \to (k_x,k_z,-k_y)$ and $(-k_z,k_y,k_x)$, respectively. It is then straightforward to show that $\hat{h}(\vec{k})$ remains invariant under O(3) rotations.

The O(3) rotational symmetry of the low-energy model is only plausible when the underlying lattice structure possesses a cubic symmetry. The isotropic Fermi velocity in a low-energy theory is then $v\sim t a$, where $t$ is the hopping amplitude and $a$ is the lattice constant. However, known Dirac materials often exhibit tetragonal symmetry~\cite{cd2as3:Exp, na3bi:Exp}, with elongated (or shortened) lattice constant along one of the three principle axes, for example. In a tetragonal environment this results in the breakdown of the O(3) rotational symmetry down to an in-plane O(2) one about the $z$-axis. In other words, $v_x=v_y \neq v_z$ in a tetragonal Dirac material. In Sec.~\ref{subsec:anisotropy} we discuss the effects of \emph{weak} rotational symmetry breaking on interacting chiral Dirac fermions in a perturbative manner.

The gapless chiral Dirac system is invariant under the discrete parity transformation ($\mathcal{P}$), reversal of time ($\mathcal{T}$), and charge conjugation ($\mathcal{C}$). In our representation of the four-component spinor $\Psi_{\vec{k}}$ and the $\Gamma$ matrices, the above transformations are realized as $\mathcal{P}\Psi_{\vec{k}} \mathcal{P}= \Gamma_{30} \Psi_{-\vec{k}}$, $\mathcal{T} \Psi^\star_{\vec{k}} \mathcal{T}= -\Gamma_{02} \Psi_{-\vec{k}}$ and $\mathcal{C} \Psi_{\vec{k}} \mathcal{C}= -i \Gamma_{12} \Psi^\star_{\vec{k}}$~\cite{peskin-schroeder}. The time-reversal operator is \emph{antiunitary} and can be written as $\mathcal{T}=U {\mathcal K}$, where $U=\Gamma_{02}$ is the unitary part and ${\mathcal K}$ is complex conjugation, such that ${\mathcal T}^2=-1$, yielding Kramers degenerate Dirac bands.

The noninteracting system remains invariant also under a continuous global U(1) chiral rotation ($\theta_{\rm ch}$), under which $\Psi_{\vec{k}} \to \exp[i \theta_{\rm ch} \Gamma_{10}] \Psi_{\vec{k}}$, where the matrix $C=\Gamma_{10}$ is the generator of the chiral rotation. We note that two mass matrices, namely $\Gamma_4$ and $\Gamma_5$, that anticommute with $\hat{h}(\vec{k})$, break the global U(1) chiral symmetry of the massless Dirac Hamiltonian. While the scalar mass ($\Gamma_4$) only breaks the continuous chiral symmetry, the pseudoscalar mass ($\Gamma_5$) in addition breaks the discrete ${\mathcal P}$ and ${\mathcal T}$ symmetries. Furthermore, the noninteracting Hamiltonian $\hat{h}(\vec{k})$ possesses a \emph{psuedospin} SU(2) chiral symmetry, which only becomes visible once we Nambu double the spinor basis, as shown in Sec.~\ref{subsec:Nambudoubing}. Therefore, the noninteracting gapless chiral Dirac system enjoys a U(1)$\otimes$SU(2) chiral symmetry.

The imaginary time ($\tau$) Euclidean action for such a collection of noninteracting chiral Dirac fermions reads
\begin{equation}
S_0=\int \D \tau \int \D^d\vec{x} \; \Psi^\dag_{\tau,\vec{x}} \left[ \partial_\tau + \hat{h}(\vec{k}\to -i\boldsymbol{\nabla} )\right]\Psi_{\tau,\vec{x}},
\end{equation}
where $d$ is the number of spatial dimensions. The absence of a chemical potential term implies the fine tuning of the Fermi energy to the band touching Dirac point, whereby the Fermi surface shrinks to just one point at the center of the Brillouin zone (see, however, Ref.~\cite{Kirkpatrick-Belitz:dopedDSM}). While constructing the interacting theory, besides the spatial rotational symmetry, we impose $\mathcal{P}$, $\mathcal{T}$ and $\mathcal{C}$ symmetries separately. But, we do not enforce invariance under U(1) chiral rotations or pseudospin SU(2) rotations explicitly. Instead, we demonstrate restoration of the chiral symmetry as an emergent phenomena at various RG fixed points that can be accessed by tuning the strength of interactions among Dirac fermions. For the time being however our noninteracting single-flavor model is chiral symmetric, which at first glance seems to be at odds with the Nielsen-Ninomiya ``no-go'' theorem~\cite{nielsen}, when it comes to constructing a corresponding lattice model. We proceed with pointing at a possible way out of this conundrum. Readers solely interested in the field theoretic results may wish to skip the following section. Nevertheless, given that two-dimensional genuine chiral Dirac-Hubbard model on a square lattice~\cite{lauchli-SLAC}, and three-dimensional Dirac-Hubbard on a cubic lattice~\cite{guo-maciejko-scaletter} have been studied recently using quantum Monte Carlo simulations, a concrete lattice realization of three-dimensional massless chiral Dirac fermions should facilitate future numerical investigations of this system, where our results can be scrutinized.


\subsection{Lattice model}~\label{subsec:latticemodel}

\begin{figure}[tbp]
\centering
\includegraphics[width=0.5\linewidth]{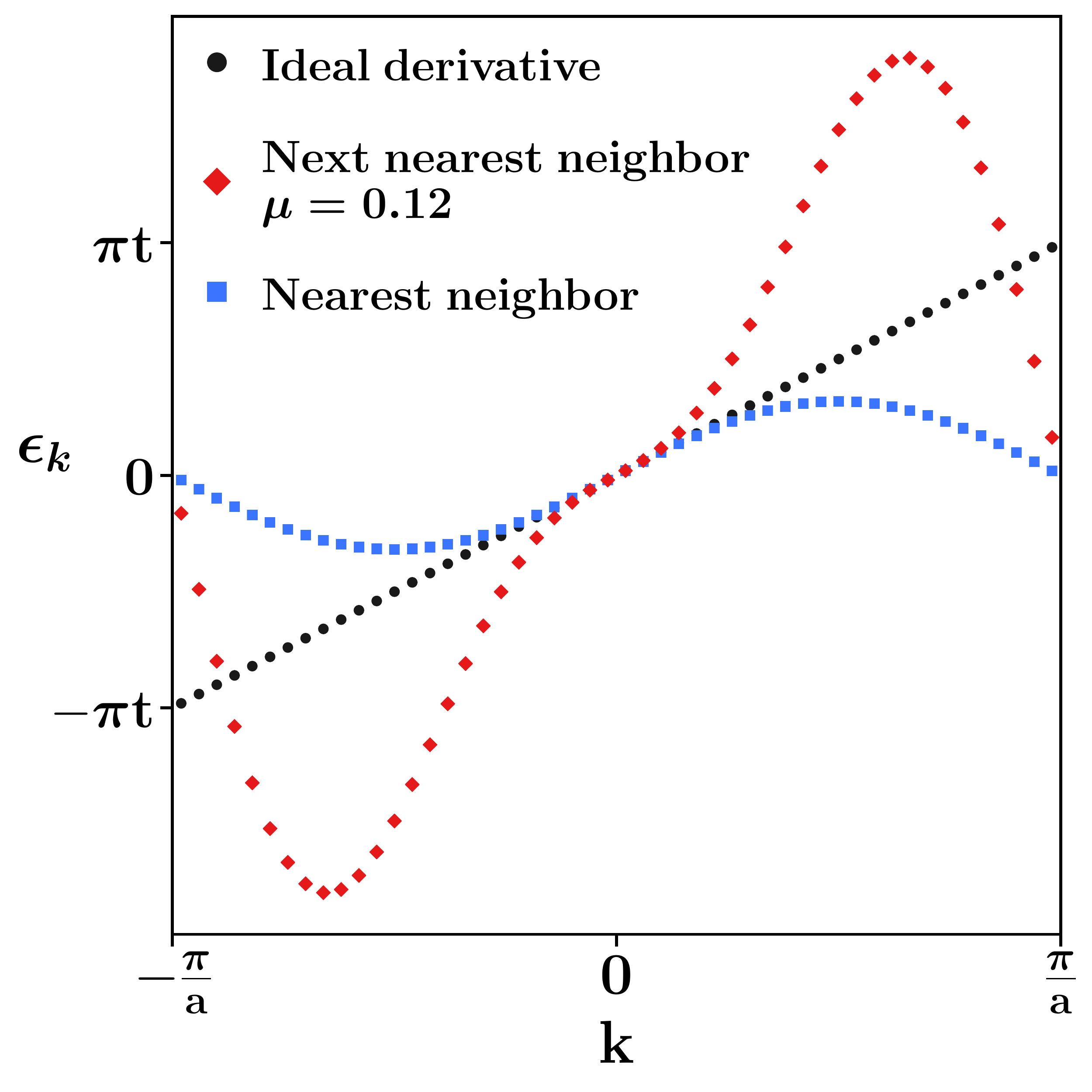}
\caption{Spectrum of the one-dimensional Hamiltonian $\hat{h}_{\rm 1D}=-i(ta) \D/(\D x)$ for three different choices for the lattice derivative. The blue squares show the simplest, nearest neighbor approach [see Eq.~(\ref{eq:nn_spectrum})], which results in doublers around $k=\pm \pi/a$. The red diamonds correspond to the next nearest neighbor spectrum from Eq.~(\ref{eq:nnn_spectrum}) for $\mu=0.12$. Notice that the Fermi velocity of the doubler around $k=\pm \pi/a$ in this case is greater than that near the center of the Brillouin zone ($k=0$). This effect is exacerbated by further decreasing $\mu$, which eventually gets rid of the doublers, but pushes much of the spectral weight to high energies ($\epsilon_k \gg \pi t$). The black dots represent the ideal derivative, the spectrum of which is devoid of doublers and remains linear in the entire Brillouin zone, thereby yielding a genuine chiral symmetry at the microscopic level. Notice that around $k \approx 0$ all three constructions recover $k$-linear dispersion. This construction can be generalized to realize three-dimensional massless Dirac fermions with microscopic chiral symmetry, see Eq.~(\ref{eq:lattice3D}). 
}~\label{fig:latticeDer}
\end{figure}

In this paper we consider a single flavor of four-component massless chiral Dirac fermions in a continuum theory, the translation of which to a lattice model is, however, a nontrivial task. The challenge originates from constructing a lattice version of the derivative operator while taking $\vec{k}\to -i\boldsymbol{\nabla}$ in Eq.~(\ref{eq:H0}). Nevertheless, symmetrizing the derivative from basic calculus as
\begin{equation}~\label{eq:nn_spectrum}
\frac{\D f(x)}{\D x}=\lim_{a\to 0} \frac{f(x+a)-f(x-a)}{2a}
\end{equation}
proffers a tempting lattice version, in which one treats $a$ as the lattice constant, and arrives at the nearest neighbor description, where in one dimension the eigenenergies are $\epsilon_k \sim t\sin(ka)$, the blue squares in Fig.~\ref{fig:latticeDer}. Here $k$ is the discrete valued lattice momentum and $t$ is a hopping amplitude, setting the energy scale. However, this construction results in so-called fermion doublers at the edges of the Brillouin zone ($k=\pm \pi/a$). In fact in $d$ dimensions one finds $2^d$ number of low-energy Dirac points compared to the continuum theory~\cite{quinn-weinstein}. A common remedy to this problem is the introduction of a momentum-dependent Wilson mass, which gaps the doublers, but vanishes at the desired $\Gamma$ point~\cite{wilson, kogut-susskind}. But, such a construction comes with its own intricacies. As such adding another discrete symmetry (${\mathcal P}$, ${\mathcal T}$, ${\mathcal C}$) preserving anticommuting mass matrix ($\Gamma_4$) spoils the genuine microscopic chiral symmetry of Dirac fermions. Even though the resulting higher order in momentum ($\propto k^2$) terms are \emph{irrelevant} in the RG framework, yielding an emergent chiral U(1) symmetry, the nucleation of ${\mathcal P}$, ${\mathcal T}$ and ${\mathcal C}$ symmetric, but chiral U(1) symmetry breaking scalar mass ($\Gamma_4$) at strong coupling then takes place through a \emph{fluctuation driven first-order transition}, as it does not break any bonafide microscopic symmetry~\cite{roy-goswami-sau}. So, we seek for a lattice realization of single flavored massless Dirac fermions with genuine microscopic chiral symmetry.

The doubler problem can also be overcome by extending the nearest neighbor derivative to next nearest neighbor~\cite{quinn-weinstein}, with hopping parameters $t_1=(1+\mu)/(2\mu)$ and $t_2=(\mu-1)/(4\mu)$ between the first and second neighbors, respectively. Here $\mu$ is a tuning parameter that recovers the nearest neighbor limit for $\mu=1$. The corresponding spectrum is then given by
\begin{equation}~\label{eq:nnn_spectrum}
\epsilon_k=t_1 \sin(k a)+t_2\sin(2 k a).
\end{equation}
Upon sending $\mu\to 0$, the Fermi velocity of the doublers gets pushed to \emph{infinity}, and due to the finite sampling of the $k$ axis there will be no additional low-energy mode in the Brillouin zone, see red diamonds in Fig.~\ref{fig:latticeDer}. However, this construction results in much of the spectral weight being pushed to high energies $\epsilon_k \gg \pi t$.

The aforementioned complications are solved by using the Stanford Linear Accelerator (SLAC) lattice derivative~\cite{nason}, a pedagogic construction of which is provided in Ref.~\cite{costella}. The three main ingredients of the SLAC construction are the following. $(i)$ Identifying the (continuum) derivative operation with convolving a function $f(x)$ with the negative derivative of the Dirac delta function $f'(x)=-\delta^\prime(x) \star f(x)$. $(ii)$ Applying a low-pass filter to the continuum formalism to restrict it in momentum space to the first Brillouin zone. $(iii)$ Finally, sampling the resulting operator on a lattice. Since the Fourier transform of the Dirac delta function is unity, and the ideal low-pass filter in position space is $\sin(\pi x/a)/(\pi x)$, steps $(i)$ and $(ii)$ can be summarized as
\begin{equation}
-(\delta^{\mathrm{filt}}(x))'=\frac{a \sin\left( \frac{\pi x}{a} \right)-\pi x \cos\left( \frac{\pi x}{a}\right)}{a \pi x^2},\label{eq:idealDer}
\end{equation}
where $\delta^{\mathrm{filt}}(x)$ is the \emph{filtered} Dirac delta operator that has Fourier components only in the first Brillouin zone. By performing the Taylor expansion we find that $\delta^{\mathrm{filt}}(x)'$ vanishes linearly around $x=0$, and the SLAC construction does not have an on site component. The ideal lattice derivative is then obtained by sampling the expression from Eq.~(\ref{eq:idealDer}) at the lattice points $x=n a$, multiplied by the lattice spacing ($a$), yielding
\begin{align}
\Delta (n)&=
\begin{cases}
   0,& \text{if } n=0\\
    - \; \frac{(-1)^n}{na},              & \text{otherwise}.
\end{cases} \label{eq:SLACDer}
\end{align}
By identifying the position space derivative with convolution from Eq.~(\ref{eq:SLACDer}), we get rid of the doublers, but pay the price of our operator now being \emph{non-local}, albeit only along the principal axes.

The resulting tight binding Hamiltonian which corresponds to Eq.~(\ref{eq:H0}) on a three-dimensional cubic lattice with linear dimension $L=2N+1$ in each direction reads
\begin{equation}~\label{eq:lattice3D}
h_\mathrm{latt}=-i\sum_{\vec{R}} \sum_{j=1}^3 \sum_{n=-N}^N  \Delta(n) \Psi^\dag_{\vec{R}} \Gamma_j \Psi_{\vec{R}+n\hat{\vec{e}}^{(j)}},
\end{equation}
where $\vec{R}$ denotes the position of the lattice sites and $\hat{\vec{e}}^{(j)}$ is a vector of length $a$ in the $j$th principal direction. While the locality of the SLAC derivative on a conceptual level can be subject to debate, the long-range nature of it is certainly unwieldy from a practical point of view. Nevertheless, in exchange we obtain spectra that reflect ``true'' momentum operator on a lattice, namely $\epsilon_{\vec{k}}=(ta) \vec{k}$, where $v=t a$ is the isotropic Fermi velocity. See the black dots in Fig.~\ref{fig:latticeDer}. As the SLAC fermion construction produces linear in momentum Dirac dispersion for the entire range of momentum $-\pi/a \leq \vec{k} \leq \pi/a$, our RG analysis is equally germane to both the continuum and lattice SLAC model for interacting massless chiral Dirac fermions, once we identify the ultraviolet momentum cutoff $\Lambda=\pi/a$, about which more in a moment.


\section{Electron-electron interactions}~\label{sec:eeinteraction}

In what follows we seek to unveil the structure of the RG fixed points (including both QCPs and BCPs) and emergent symmetries therein starting from an interacting model for three-dimensional massless chiral Dirac fermions. To capture interaction-induced spontaneous symmetry breaking in this system we construct the symmetry-allowed four-fermion or quartic terms. In this paper we only take the Hubbardlike short-range (momentum-independent) interactions into account. On the other hand, the long-range tail of the Coulomb interaction in Dirac materials only provide logarithmic correction to the Fermi velocity~\cite{goswami-chakravarty, isobe-nagaosa, jose-gonzalez, prokofev, dassarma-throckmorton, juricic-herbut-semenoff, drut-lahde, roy-dassarma, katslenson, zhao-wang:Coulomb, yang-wang-liu:Coulomb}, without causing any transition to ordered states~\cite{juricic-herbut-semenoff, roy-dassarma, prokofev}. When simultaneously present with the short-range interactions, it can only cause non-universal shifts of the phase boundaries~\cite{drut-lahde, katslenson}, without altering the underlying quantum critical behavior~\cite{juricic-herbut-semenoff, roy-dassarma}.

The most general local four-fermion term is of the schematic form
\begin{equation}
g_{_{\mu \nu \rho \lambda}} (\Psi^\dag \Gamma_{\mu \nu} \Psi)(\Psi^\dag \Gamma_{\rho \lambda} \Psi), \nonumber 
\end{equation}
where $\mu,\nu,\rho,\lambda=0,1,2,3$, $g_{_{\mu \nu \rho \lambda}}$ is the coupling constant, and $\Psi^\dag \equiv \Psi^\dag_{\tau,\vec{x}}$ and $\Psi \equiv \Psi_{\tau,\vec{x}}$ are two independent Grassmann variables in the path integral or action formalism. Before imposing any symmetry constraint on the quartic terms, altogether there are 136 of them. Namely, 16 (the number of Hermitian matrices, constituting the basis for all four-dimensional matrices) of them are obtained for $\mu \nu=\rho \lambda$, while $\mu \nu \neq \rho \lambda$ in the remaining 120 quartic terms. However, this number gets drastically reduced first by imposing the discrete symmetries, namely parity (${\mathcal P}$), time-reversal (${\mathcal T}$) and charge conjugation (${\mathcal C}$). These three discrete symmetries permit only 12 terms with $\mu \nu \neq \rho \lambda$ and 16 quartic terms with $\mu \nu=\rho \lambda$. The spatial O(3) rotational symmetry eliminates the former set of quartic terms, and organizes remaining 16 terms of the form $(\Psi^\dag \Gamma_{\mu \nu} \Psi)^2$ into eight distinct interaction channels. For a detailed derivation see Appendix~\ref{seq:bilinears_class}. The interacting Lagrangian containing all symmetry allowed local quartic terms is then given by
\begin{equation}~\label{eq:Lint}
L_{\mathrm{int}}= \sum^3_{\mu=0} g^s_{_\mu} (\Psi^\dag \Gamma_{\mu 0} \Psi)^2
+ \sum^3_{\mu=0} g^t_{_\mu} \left[ \sum^3_{j=1}(\Psi^\dag \Gamma_{\mu j} \Psi)^2 \right].
\end{equation}
The four-fermion interactions with coupling constants $g^{s}_{_j}$ ($g^{t}_{_j}$) for $j=1,2,3$ correspond to spin-independent (spin-dependent) mixing of even and odd parity states. By contrast, $g^s_{_0}$ ($g^t_{_0}$) corresponds to short-range density-density (ferromagnetic) interaction. Notice that we did not enforce chiral U(1) symmetry on $L_{\mathrm{int}}$ at the bare level. Consequently, the components of chiral U(1) vectors appear as independent quartic terms in Eq.~(\ref{eq:Lint}). Namely, two terms containing $\Gamma_{20}$ and $\Gamma_{30}$ ($\Gamma_{2j}$ and $\Gamma_{3j}$) acquire two separate coupling constants $g^s_{_2}$ and $g^s_{_3}$ ($g^t_{_2}$ and $g^t_{_3}$), respectively, at the microscopic level. On other hand, in Weyl semimetals the chiral U(1) symmetry is associated with the translational symmetry in the continuum limit~\cite{maciejko-nandkishore:Weyl, roy-goswami-juricic:Weyl}. Therefore, the interacting theory must preserve the chiral U(1) symmetry in Weyl semimetals.

However, the number of linearly independent four-fermion terms is only \emph{four} due to the existence of the Fierz identity~\cite{herbut-juricic-roy} (see Appendix~\ref{sec:Fierz}). We choose them to be the ones appearing with the coupling constants $\{ g^s_{_\mu}\}$. The corresponding interacting Euclidean action reads
\begin{equation}~\label{eq:S_int}
S_{\mathrm{int}}=\int \D \tau \int \D^d \vec{x} \; \left(\; \sum^3_{\mu=0} g^s_{_\mu} \; (\Psi^\dag \Gamma_{\mu 0} \Psi)^2 \right).
\end{equation}
Notice that eight quartic terms appearing in $L_{\mathrm{int}}$ can be decomposed into two sectors, the ones transforming as scalars (three-component vectors) under spatial O(3) rotations and appearing with coupling constants $g^s_{_\mu}$ ($g^t_{_\mu}$), for $\mu=0, \cdots, 3$. Consequently, we can choose four quartic terms in the singlet channel as the independent ones. Next we study the interacting model $S_0+S_{\mathrm{int}}$ within the framework of Wilsonian momentum-shell RG analysis.


\subsection{Renormalization group analysis}

To study the quantum critical properties of this system, we examine its low-energy and long-wavelength behavior under the tuning of non-thermal parameters, such as the strength of the local interactions in this case. To proceed we introduce a hard ultraviolet momentum cutoff $\Lambda = \pi/a$, which replaces the cubic Brillouin zone by a spherical one (see Sec.~\ref{subsec:latticemodel} for a lattice origin of $\Lambda$ for SLAC massless chiral Dirac fermions). Here $a$ bears the dimension of the lattice constant. In the RG procedure we then gradually decrease $\Lambda$ by repeatedly integrating out a thin Wilsonian momentum shell defined as $\Lambda e^{-\ell}<|\vec{k}|<\Lambda$, where $\ell > 0$ is the logarithm of the RG scale. Finally, we restore the Euclidean action ($S_0+S_{\mathrm{int}}$) into its original form, but in terms of the renormalized quantities.

The scaling of various quantities appearing in the action relative to $\vec{k}$ is of crucial importance as we gradually lower the ultraviolet cutoff. The scaling dimensions of momentum and frequency are respectively $\left[ k \right] = 1$ and $\left[ \omega \right] = z$, where the dynamic scaling exponent $z=1$ for linearly dispersing Dirac fermions. The scale-invariance of $S_0$ requires that $\left[ \Psi \right] = d/2$, from which we obtain the scaling dimension of the quartic coupling constants to be $[ g^s_{_\mu}] = z-d$. Therefore, local interactions in three-dimensional Dirac systems are \emph{irrelevant} in the RG sense and the Dirac cone remains stable as long as they are sufficiently weak. On the other hand, strong enough local interactions can drive the system through quantum phase transitions into various broken symmetry phases. The scaling dimension $[ g^s_{_\mu} ]$ pins the lower critical dimension at $d=1$, where short-range interactions are marginal, which facilitates a controlled $\epsilon$ expansion about one spatial dimension, with $\epsilon=d-1$. Note that vanishing density of states, namely $\rho(E)\sim |E|^2$ indicates stability of the Dirac node against sufficiently weak interactions in $d=3$, whereas interactions become marginal when $d=z=1$, yielding a constant density of states.

Evaluating the relevant Feynmann diagrams up to one-loop order~\cite{szabo-moessner-roy}, we arrive at the following leading-order RG flow equations for the coupling constants
\allowdisplaybreaks[4]
\begin{align}~\label{eq:beta_g}
\frac{\D g^s_{_0}}{\D \ell}=&-\epsilon g^s_{_0} - \left( g^s_{_0} g^s_{_3} + g^s_{_0} g^s_{_2} + 2 g^s_{_3} g^s_{_2} \right),\nonumber \\
\frac{\D g^s_{_1}}{\D \ell}=&-\epsilon g^s_{_1} + g^s_{_0} g^s_{_3} + g^s_{_0} g^s_{_2} - 4 g^s_{_3} g^s_{_2}, \nonumber \\
\frac{\D g^s_{_2}}{\D \ell}=&-\epsilon g^s_{_2} + 3 \left( g^s_{_2} \right)^2 -2 \left( g^s_{_0} g^s_{_3} + g^s_{_0} g^s_{_2} - g^s_{_3} g^s_{_2} \right)+3 g^s_{_1} \left( g^s_{_2} - g^s_{_3} \right),\nonumber \\
\frac{\D g^s_{_3}}{\D \ell}=&-\epsilon g^s_{_3} + 3 \left( g^s_{_3} \right)^2 -2 \left( g^s_{_0} g^s_{_3} + g^s_{_0} g^s_{_2} - g^s_{_3} g^s_{_2} \right)+3 g^s_{_1} \left( g^s_{_3} - g^s_{_2} \right).
\end{align}
Here we made the substitution $\frac{\Lambda^\epsilon}{3 v \pi^2} g^s_{_\mu} \to g^s_{_\mu}$, such that the flow equations are expressed in terms of dimensionless couplings. Note that the flow equations are symmetric under the exchange of $g^s_{_2}$ and $g^s_{_3}$, manifesting the chiral U(1) symmetry of the noninteracting system. Here all the matrix algebra are performed in $d=3$ and subsequently the radial integral in the momentum space is carried out in dimension $d=1+\epsilon$.

\subsection{Fixed points and critical exponents}

We examine the quantum critical phenomena by analyzing the fixed point structure of the RG flow equations from Eq.~(\ref{eq:beta_g}), which from now we denote as $\beta$ functions $\beta_{g^s_\mu}\equiv \D g^s_\mu/\D \ell$. A fixed point $\vec{g}^s_\ast$ in the four-dimensional coupling constant space is found by simultaneously solving $\beta_{g^s_\mu}(\vec{g}^s_{_\ast})=0$, where $\mu=0,\cdots,3$ and $\vec{g}^s=(g^s_{_0},g^s_{_1},g^s_{_2},g^s_{_3})$. Each fixed point is characterized by linearizing the RG flow around it. To this end we construct a four-dimensional stability matrix ($\vec{M}$) and compute its eigenvalues at various fixed points. The elements of the stability matrix are explicitly given by  
\begin{equation}
M_{\mu+1,\nu+1}=\frac{\D \beta_{g^s_\mu}}{\D g^s_\nu}, 
\end{equation}
where $\mu,\nu=0,\cdots, 3$. A negative (positive) eigenvalue of $\vec{M}$ corresponds to a stable (unstable) eigendirection, and the number of unstable directions characterizes a given fixed point. For example, a QCP (BCP) has one (two) unstable direction(s).

We find all together nine fixed points~\cite{roy-dassarma}, of which the trivial one at $\vec{g}^s=(0,0,0,0)$ is the fully attractive noninteracting Gaussian fixed point. It describes a stable Dirac semimetal for sufficiently weak, but generic short-range interactions. The rest of the eight fixed points at nontrivial strength of the coupling constants are reported in Table~\ref{tab:FPs}. Four of them are QCPs (C1,$\cdots$, C4) and the remaining four are BCPs (B1, $\cdots$, B4). The existence of BCPs is necessary for the continuity of the RG flow trajectories. As such they separate the basins of attraction of various QCPs. All fixed points are located at $\vec{g}^s_{_\ast} \sim \epsilon$, which can be seen from Eq.~(\ref{eq:beta_g}).

The dynamic scaling exponent $z=1$ at all four QCPs. On the other hand, the inverse of the positive eigenvalue of the stability matrix ($\vec{M}$) determines the correlation length exponent ($\nu$) at each QCP. To the leading order in the $\epsilon$ expansion we obtain $\nu^{-1}=\epsilon$, and $\nu=1/2$ in $d=3$. The mean-field value of the correlation length exponent is an exact result, as the system resides at the upper-critical three spatial dimensions~\cite{zinn-justin:book, zinn-justin-moshe-moshe}.

From the locations of the fixed points we can conclude the following. Two QCPs C1 and C4 are chiral symmetric, where the couplings of the two components of the U(1) chiral vector $g^s_{_2}$ and $g^s_{_3}$ are \emph{identical}. The remaining two QCPs, namely C2 and C3, are chiral partners of each other and their locations transform into one another under the chiral rotation $g^s_{_2} \leftrightarrow g^s_{_3}$. On the other hand, the four BCPs form two such pairs of chiral partners, namely under $g^s_{_2} \leftrightarrow g^s_{_3}$ (1) B1 $\leftrightarrow$ B2 and (2) B3 $\leftrightarrow$ B4. To further anchor the restoration of chiral symmetry at various RG fixed points next we compute the scaling dimension of all symmetry allowed fermion bilinears.

\begin{figure}[tbp]
\includegraphics[width=0.48\linewidth,height=4.5cm]{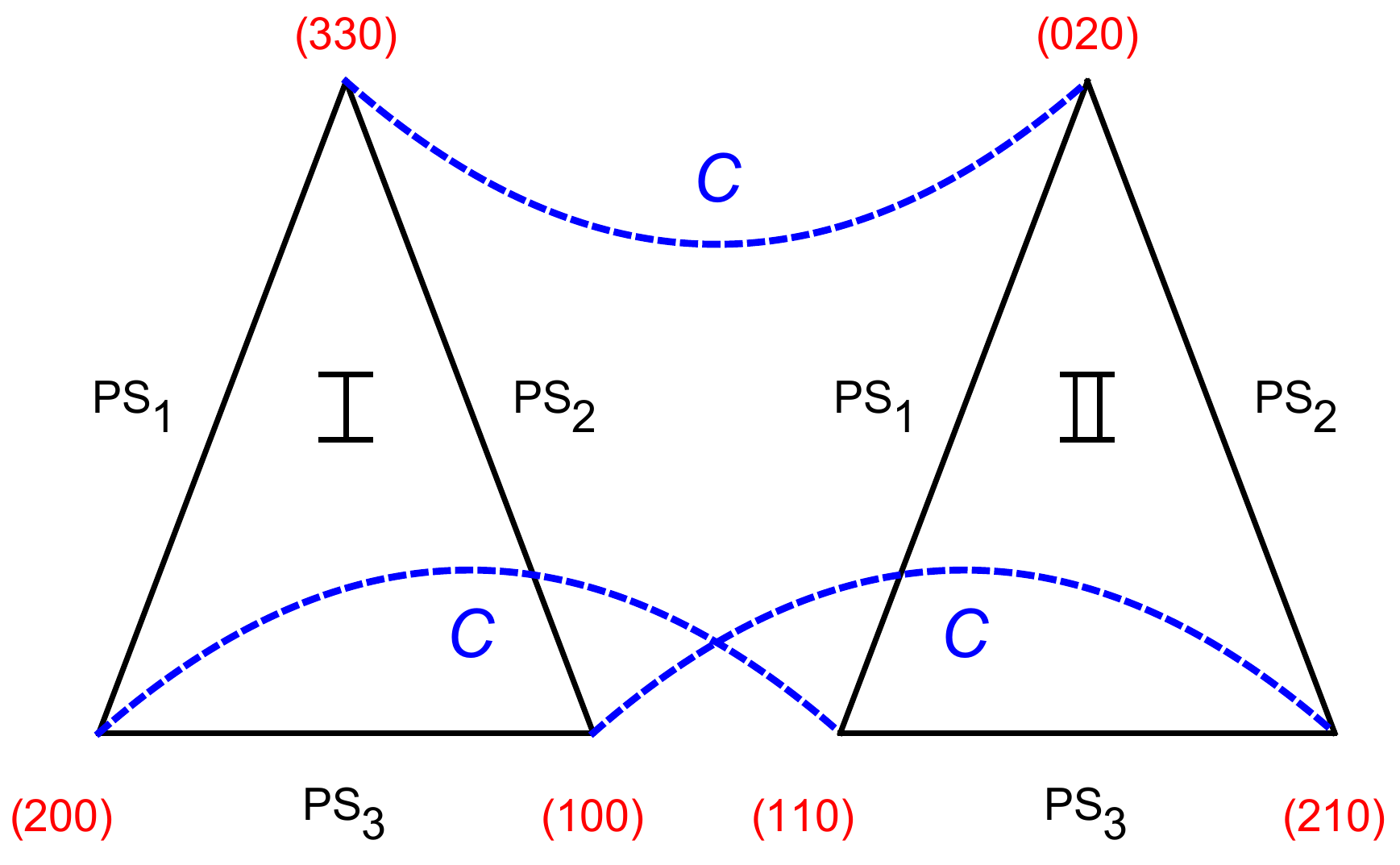}
\includegraphics[width=0.48\linewidth,height=4.5cm]{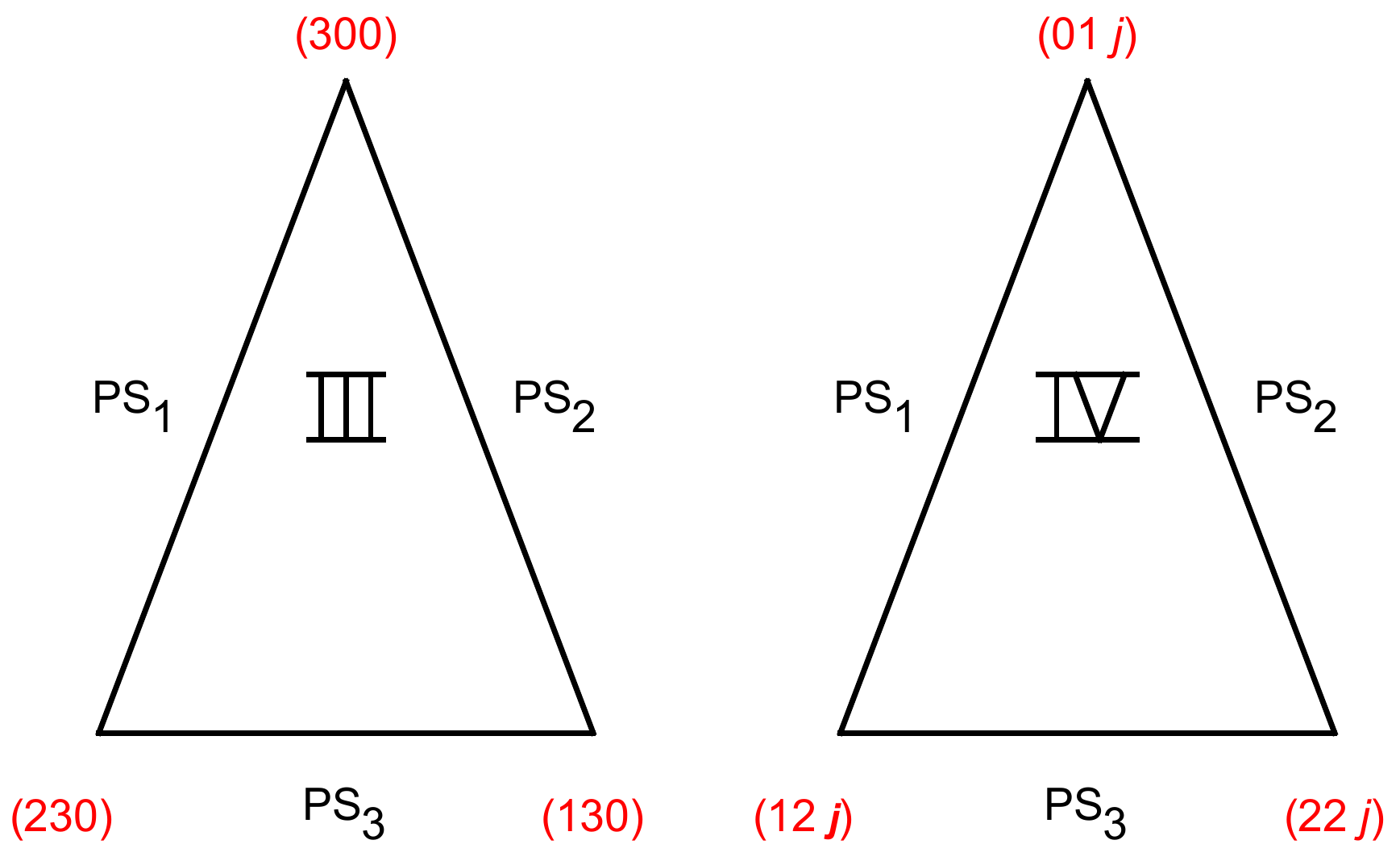}
\caption{Four sets of three mutually anticommuting fermion bilinears in the Nambu doubled basis $\Psi_{\rm Nam}$ [see Eq.~(\ref{eq:Nambudefinition})], denoted by $(\mu \nu \rho) \equiv \Psi^\dag_{\rm Nam} \eta_\mu \Gamma_{\nu \rho} \Psi_{\rm Nam}$, residing at the vertices of four triangles, each representing a pseudospin SU(2) vector. Each triangle demonstrates pseudospin SU(2) chiral rotation among (I) scalar, (II) pseudoscalar excitonic and two components (real and imaginary) of pairing or superconducting masses, (III) fermionic density and two components of temporal vector pairing, and (IV) Abelian current and two components of spatial vector pairing with a specific spin orientation ($\rho \equiv j=1,2,3$). The real and imaginary components of any pairing correspond to $\mu=1$ and $2$, respectively. Three pseudospin generators PS$_j$ with $j=1,2,3$ reside at three vertices of III [see also Eq.~(\ref{eq:pseudospingenerator})]. Any arm of the triangle corresponds to the rotation between two fermion bilinears, sitting at its two ends, by a specific generator of pseudospin SU(2) chiral symmetry. Also the identical vertices of triangles I and II are related by the chiral U(1) rotation (the blue dashed lines), generated by $C=\eta_3\Gamma_{10}$. See also Table~\ref{tab:bilinears}.    
}~\label{fig:pseudospin}
\end{figure}

\subsection{Nambu doubling and emergent chiral symmetry}~\label{subsec:Nambudoubing}

Sufficiently strong local electronic interactions destabilize a gapless Dirac liquid and drive the system through quantum phase transitions into various broken symmetry phases. Here we consider all such symmetry allowed particle-hole (or excitonic) and particle-particle (or superconducting or pairing) orders. To bring both sectors under a unified framework we extend the Dirac spinor following the Nambu doubling
\begin{equation}~\label{eq:Nambudefinition}
\Psi_{\vec{k}} \to \Psi_{\rm Nam}=
\left[
\begin{array}{c}
\Psi_{\vec{k}} \\
\Gamma_{20} \Psi^\ast_{-\vec{k}}
\end{array}\right].
\end{equation} 
Note that in the lower block we absorb the unitary part of the time-reversal operator (${\mathcal T}$), such that $\Psi_{\rm Nam}$ transform as $\Psi_{\vec{k}}$ under all symmetry operations~\cite{roy-ghorashi-foster-nevidomskyy}. The noninteracting Hamiltonian from Eq.~(\ref{eq:H0}) in this basis reads 
\begin{equation}~\label{eq:H0Nambu}
\hat{h}_{\rm Nam}(\vec{k})= \eta_3 \: \sum^{3}_{j=1} v_j\Gamma_{j}k_j. 
\end{equation}
The newly introduced set of Pauli matrices $\{ \eta_\mu \}$ operate on the Nambu or particle-hole indices, with $\mu=0, \cdots, 3$. The generator of the chiral U(1) symmetry in this basis becomes $C=\eta_3 \Gamma_{10}$. Furthermore, the Nambu doubling allows us to unveil the \emph{psedospin} SU(2) chiral symmetry of the noninteracting system, generated by 
\begin{equation}~\label{eq:pseudospingenerator}
\vec{{\rm PS}}=\left\{ \eta_1 \Gamma_{30}, \eta_2 \Gamma_{30},\eta_3 \Gamma_{00} \right\},
\end{equation}
since $[\hat{h}_{\rm Nam}(\vec{k}), \vec{{\rm PS}}]=0$.~Specifically, ${\rm PS}_3=\eta_3 \Gamma_{00}$ is the number operator.~Therefore a collection of noninteracting four-component massless Dirac fermions enjoys U(1)$\otimes$SU(2) chiral symmetry, since $[C,\vec{{\rm PS}}]=0$.~Previously, the pseudospin SU(2) symmetry has been discussed in the context of two-dimensional Hubbard model~\cite{roy-foster:PRX, pseudospin-1, pseudospin-2, pseudospin-3}.~But, its imprints on three-dimensional interacting systems remained unexplored so far.

To appreciate the imprint of the enlarged chiral symmetry in the presence of interactions, next we consider all symmetry allowed fermion bilinears describing different orders in the Nambu doubled basis. The effective action containing all (excitonic and pairing) symmetry allowed local orders reads as
\begin{equation}~\label{eq:h_susc}
S_{\mathrm{local}}=\int \D \tau \int \D^d \vec{r} \Psi_{\rm Nam}^\dag (\hat{h}_{\mathrm{exc}}+\hat{h}_{\mathrm{pair}}) \Psi_{\rm Nam},
\end{equation}
where
\allowdisplaybreaks[4]
\begin{align}
\hat{h}_{\mathrm{exc}}&=\Delta_0^s \eta_3\Gamma_{00} + \Delta_1^s \eta_3\Gamma_{10} +\Delta_2^s \eta_0\Gamma_{20}  + \Delta_3^s \eta_3\Gamma_{30} \nonumber\\
&+\sum_{j=1}^3 \Bigg[ \Delta_0^t \eta_0\Gamma_{0j} +\Delta_1^t \eta_0\Gamma_{1j} + \Delta_2^t \eta_3\Gamma_{2j} + \Delta_3^t \eta_0\Gamma_{3j} \Bigg],
\end{align}
and
\allowdisplaybreaks[4]
\begin{align}~\label{eq:pairingsingleparticle}
\hat{h}_{\mathrm{pair}}&=(\eta_1 \cos \phi + \eta_2 \sin \phi) 
\times \Big[ \Delta_0^p \Gamma_{00} + \Delta_1^p \Gamma_{10}
+ \Delta_2^p\sum_{j=1}^3 \Gamma_{2j} + \Delta_3^p \Gamma_{30}  \Big],
\end{align}
with $\phi$ as the U(1) superconducting phase. The real (imaginary) component of any pairing order corresponds to $\phi=0$ ($\pi/2$). Here $\Delta^a_\mu$ is the conjugate field of the corresponding fermion bilinear.

The bilinears, their physical meanings together with the corresponding matrices and their transformations under various symmetry operations (discrete and continuous) are summarized in Table~\ref{tab:bilinears}. The two fully gapped phases in the particle-hole subspace are the scalar and pseudoscalar masses, which form a U(1) vector under the chiral U(1) rotation generated by $C=\eta_3 \Gamma_{10}$. The Nambu basis accommodates two additional massive orders in the particle-particle sector, the scalar $s$-wave and pseudoscalar $p$-wave pairings~\cite{ohsaku, fu-berg}. They form two copies of U(1) chiral vector, where the doubling is due to the gauge redundancy in the internal U(1) degree of freedom associated with the superconducting phase ($\phi$). Furthermore, the two tensor bilinears (coupled with the conjugate fields $\Delta^t_2$ and $\Delta^t_3$) form other three copies of two-component chiral U(1) vector, where the three-fold redundancy stems from the spatial O(3) symmetry. Additionally, there exist four three-component orders, each of which is a composite of a particle-hole and a particle-particle order and transforms as a vector under the pseudospin SU(2) chiral rotations, generated by $\vec{{\rm PS}}$ [see Eq.~(\ref{eq:pseudospingenerator})], as shown in Fig.~\ref{fig:pseudospin}.

To extract the scaling dimension of various fermion bilinears, first we compute the RG flow equations for the corresponding source terms or conjugate fields $\Delta^a_\mu$, where $\mu=0, \cdots, 3$ and $a=s,t,p$. After evaluating the relevant Feynmann diagrams up to the one-loop order~\cite{szabo-moessner-roy}, we arrive at the following leading-order $\beta$ functions for $\Delta^a_\mu$
\allowdisplaybreaks[4]
\begin{align}~\label{eq:beta_Delta}
\bar{\beta}_{\Delta_0^s} &= 0, & \bar{\beta}_{\Delta_1^s} &= 0,\nonumber \\
\bar{\beta}_{\Delta_2^s} &= -\frac{3}{2} (g^s_{_0} - g^s_{_1} - 3 g^s_{_2} - g^s_{_3}), &
\bar{\beta}_{\Delta_3^s} &= -\frac{3}{2}(g^s_{_0}-g^s_{_1}-g^s_{_2}-3g^s_{_3}), \nonumber \\
\bar{\beta}_{\Delta_0^t} &= -g^s_{_0} - g^s_{_1} - g^s_{_2} - g^s_{_3}, &
\bar{\beta}_{\Delta_1^t} &= -g^s_{_0} - g^s_{_1} + g^s_{_2} + g^s_{_3}, \nonumber \\
\bar{\beta}_{\Delta_2^t} &= \frac{1}{2}(-g^s_{_0} + g^s_{_1} - g^s_{_2} + g^s_{_3}), &
\bar{\beta}_{\Delta_3^t} &= \frac{1}{2} (-g^s_{_0} + g^s_{_1} + g^s_{_2} - g^s_{_3}), \nonumber \\
\bar{\beta}_{\Delta_0^p} &= \frac{3}{2} (g^s_{_0} + g^s_{_1} - g^s_{_2} + g^s_{_3}), & 
\bar{\beta}_{\Delta_1^p} &= \frac{3}{2} (g^s_{_0} + g^s_{_1} + g^s_{_2} - g^s_{_3}), \nonumber \\
\bar{\beta}_{\Delta_2^p} &= g^s_{_0} - g^s_{_1} - g^s_{_2} - g^s_{_3}, & \bar{\beta}_{\Delta_3^p} &= 0,
\end{align}
where $\bar{\beta}_{\Delta^a_\mu} = \D \ln \Delta^a_\mu/\D \ell-1$. The right hand side of each equation corresponds to the one-loop corrections to the scaling dimension of the corresponding order or fermion bilinear, which we compute at various RG fixed points (see Table~\ref{tab:FPs}) and their values are reported in Table~\ref{tab:AnomDim}.

Computation of one-loop corrections to the scaling dimensions for all excitonic and pairing orders reveals the emergent chiral symmetry at various RG fixed points. First we note that the bilinears that commute with $\hat{h}_{\rm Nam}(\vec{k})$, the fermionic density, chiral density and temporal vector pairing, do not receive any correction to their bare scaling dimension. The matrices associated with these bilinears are also the generators of U(1)$\otimes$SU(2) chiral symmetry of the noninteracting system.

One can construct three chiral U(1) vectors by combining (1) the scalar and pseudoscalar excitonic masses, (2) two tensor order parameters, and (3) the scalar $s$-wave and pseudoscalar $p$-wave pairing masses, see Table~\ref{tab:bilinears}. Two components of any chiral U(1) vector acquire identical scaling dimensions at C1 and C4, while their scaling dimensions are exchanged between (a) C2 and C3, (b) B1 and B2, and (c) B3 and B4. Therefore, only C1 and C4 are chiral U(1) symmetric. On the other hand, the four-dimensional coupling constant space accommodates three chiral U(1) partner fixed points, namely (a) (C2,C3), (b) (B1,B2), and (c) (B3,B4), as we previously anticipated from their locations.

Also the possible restoration of the pseudospin SU(2) chiral symmetry can be anchored from the scaling dimensions of fermion bilinears at various RG fixed points. As such one can construct four three-component pseudospin SU(2) vectors by combining (1) the scalar excitonic and pairing masses, (2) the pseudoscalar excitonic and paring masses, (3) fermion density and temporal vector pairing, and (4) Abelian current and spatial vector pairing, see Fig.~\ref{fig:pseudospin}. All components of each pseudospin vector acquire identical scaling dimensions at three QCPs, C2, C3 and C4, and at two BCPs, B3 and B4. Hence, these five fixed points are pseudospin SU(2) chiral symmetric. Therefore, while all the RG fixed points at least partially restore the chiral U(1)$\otimes$SU(2) symmetry of the noninteracting system, it is fully restored only at the C4 QCP.

Even though the space of independent coupling constants is four-dimensional, the imprint of various chiral symmetric fixed points on the global phase diagram of interacting massless chiral Dirac fermions can be appreciated by focusing on its representative cut on the $(g^s_{_3},g^s_{_2})$ plane, as shown in Fig.~\ref{fig:phasediag}, which we discuss next.


\subsection{Phase diagram}

Now we proceed to construct the phase diagram in the $(g^s_{_3},g^s_{_2})$ plane by numerically solving the $\beta$ functions of the coupling constants [see Eq.~(\ref{eq:beta_g})] and the conjugate order parameter fields [see Eq.~(\ref{eq:beta_Delta})]. We simultaneously run the flow equations for four coupling constants ($g^s_\mu$) and all symmetry allowed source terms ($\Delta^a_\mu$) for various choices of the \emph{bare} coupling constants as a function of the RG time ($\ell$). Sufficiently weak but generic local interactions are irrelevant perturbation in three-dimensional Dirac semimetals due to the vanishing density of states ($\rho(E) \sim |E|^2$). Therefore, onset of any ordered phase takes place beyond a bare critical strength of interaction, which we identify from the divergence of at least one of the renormalized coupling constants as $\ell \to \infty$. At the same time at least one of the source terms diverges, which ultimately determines the pattern of the symmetry breaking in the ordered phase in an unbiased fashion. We pursue this approach to construct a representative cut of the global phase diagram for three-dimensional interacting massless chiral Dirac fermions, displayed in Fig.~\ref{fig:phasediag}, which manifests an intriguing confluence of all \emph{four} mass orders (the scalar and pseudoscalar excitonic and pairing masses), and the restoration of chiral and pseudospin symmetries among them. Note that at zero temperature nucleation of mass orders is energetically most favored as they isotropically gap the Dirac point and thereby optimally lower the free-energy (no competition with entropy~\cite{szabo-moessner-roy}). Next we highlight the role of various RG fixed points on different segments of this phase diagram.

The chiral U(1) symmetric QCP C1 can be accessed for \emph{purely} repulsive interactions $g^s_{_2},g^s_{_3}>0$ and it governs the continuous quantum phase transitions between Dirac semimetal and the two excitonic mass orders (the blue phase boundary in Fig.~\ref{fig:phasediag}). These two excitonic masses are degenerate along the 45$^\circ$ red line, where we realize an axionic insulator~\cite{peccei-quinn, weinberg, wilczek}. A slight deviation from the red line in favor of $g^s_{_3}$ ($g^s_{_2}$) tips the balance in favor of the nucleation of pure scalar (pseudoscalar) excitonic mass.

The phase transitions out of the Dirac semimetal across the black phase boundary into the pseudoscalar excitonic and pairing ($p$-wave superconductor) masses is governed by the SU(2) pseudospin symmetric QCP C2. In the absence of a Fermi surface, strong attractive interaction in at least one channel is required for the stability of a superconducting phase, which in this case occurs in the $g^s_{_3}$ channel. These two orderings are degenerate along the 135$^\circ$ (orange) line. The basins of attraction of C1 and C2 are separated by the bicritical point B2, where the pseudoscalar exciton possesses the largest scaling dimension, see Table~\ref{tab:AnomDim}. Still the transition to this ordered phase through B2 is continuous as it is accessed by holding one of its unstable directions fixed~\cite{roy-foster:PRX}.

The rest of the phase diagram and the role of various fixed points therein can be appreciated by exploiting a mirror symmetry of the whole phase diagram, under which $g^s_{_2} \leftrightarrow g^s_{_3}$. Such a mirror symmetry is rooted in the underlying U(1) chiral symmetry and corresponds to a reflection about the 45$^\circ$ diagonal (dashed) line, defined by $g^s_{_2}=g^s_{_3}$, that brings the upper left triangle of the phase diagram onto its chiral partner, the bottom right triangle. Also note that the three-component scalar and pseudoscalar masses, constructed by combining the corresponding excitonic and pairing orders, can be rotated into each other by the generator of the chiral U(1) symmetry $C=\eta_3 \Gamma_{10}$, see Fig.~\ref{fig:pseudospin}. Correspondingly, C2 is rotated into C3, which governs the phase transitions between the Dirac semimetal and the scalar excitonic and pairing ($s$-wave superconductor) mass orders that take place through the purple phase boundary. These two competing orders are degenerate along the 135$^\circ$ (pink) line, as C3 is bestowed with the same pseudospin SU(2) chiral symmetry. In this segment of the phase diagram, the roles of $g^s_{_2}$ and $g^s_{_3}$ are exchanged due to the aforementioned mirror transformation and the requisite strong attractive interaction for the nucleation of the scalar $s$-wave pairing is now fulfilled by $g^s_{_2}(<0)$. The basins of attraction of C1 and C3 are separated by B1, where the scalar excitonic mass possesses the largest scaling dimension, see Table~\ref{tab:AnomDim}. Still the transition to this phase through B1 is continuous.

The remaining three fixed points, C4, B1 and B2, play no evident role on the phase diagram in the $(g^s_{_3},g^s_{_2})$ plane. Moreover, we find that the existence of the QCP C4 is due to the assumed O(3) rotational symmetry and is in general not present in a tetragonal environment. Next we address the effects of \emph{weak} rotational symmetry breaking on various RG fixed points in a perturbative manner. \\

\subsection{Anisotropic Dirac semimetal}~\label{subsec:anisotropy}

Finally, we address the breakdown of the O(3) rotational symmetry down to an in-plane O(2) invariance about the $z$-axis, which manifests through an anisotropy between the perpendicular and the $z$ components of the Fermi velocity, i.e., $v_x=v_y=v_\perp \neq |v_z|$. Due to such a reduced symmetry, all three-component four-fermion terms get split into the perpendicular and $z$ components, namely $g^t_{\mu} \to (g^\perp_{\mu}, g^z_{\mu})$, where $\mu=0, \cdots, 3$, see Eq.~(\ref{eq:Lint}). The interacting Lagrangian then contains 12 quartic terms, out of which only \emph{five} are linearly independent due to the Fierz constraint. See Appendix~\ref{sec:Fierz} for detailed derivation. To account for the reduced symmetry, we take the fifth independent quartic term to be $g^z_{_1} (\Psi^\dag \Gamma_{13} \Psi)^2$. Assuming sufficiently \emph{weak} rotational symmetry breaking, we compute the one-loop renormalization of $g^z_{_1}$ only \emph{up to linear order} in $g^z_{_1}$, yielding
\begin{equation}~\label{eq:beta_g8z}
\frac{\D g^z_{_1}}{\D \ell}= \left(-\epsilon -g^s_{_0}-3 g^s_{_1}-2 g^s_{_2}-2 g^s_{_3}  \right) g^z_{_1} + \mathcal{O}\left( (g^z_{_1})^2\right).
\end{equation}
Since we are interested in capturing the leading order effects of \emph{weak} rotational symmetry breaking, all the Feynman diagrams are computed with isotropic Dirac kernel.

The relevance of the rotational symmetry breaking can be estimated by computing the scaling dimension of $g^z_{_1}$ at all the fixed points reported in Table~\ref{tab:FPs} for the isotropic system. Evaluating the right hand side of Eq.~(\ref{eq:beta_g8z}) at the eight fixed points from Table~\ref{tab:FPs}, we find
\allowdisplaybreaks[4]
\begin{align}~\label{eq:bg8_QCPs}
&\left. \frac{\D \ln g^z_{_1}}{\D \ell}\right\rvert_{\mathrm{C1}}= -1.251 \epsilon, &
&\left. \frac{\D \ln g^z_{_1}}{\D \ell}\right\rvert_{\mathrm{C2/C3}}= -1.594\epsilon, &
&\left. \frac{\D \ln g^z_{_1}}{\D \ell}\right\rvert_{\mathrm{C4}}= \epsilon, \nonumber\\
&\left. \frac{\D \ln g^z_{_1}}{\D \ell}\right\rvert_{\mathrm{B1/B2}}= -\frac{5}{3} \epsilon, &
&\left. \frac{\D \ln g^z_{_1}}{\D \ell}\right\rvert_{\mathrm{B3/B4}}= -0.155\epsilon.  & &
\end{align}
Therefore, slight distortion of the Dirac cone is an \emph{irrelevant} perturbation in the close vicinity of all the fixed points, except C4. In other words, C4 turns into a BCP even for sufficiently \emph{weak} breaking of the rotational symmetry. Therefore, this QCP cannot be found in general in a three-dimensional interacting Dirac system. We also note that the irrelevance of $g^z_{_1}$ at the BCPs B3 and B4 is \emph{weak}, in comparison to those near C1, C2, C3, B1 and B2. Therefore, it is conceivable that these two BCPs ultimately turn into tricritical points with three relevant directions in a strongly anisotropic Dirac semimetal, which we leave for a future investigation.

\section{Summary and Discussion}~\label{sec:summary}

Here we investigate the role of strong momentum-independent local or Hubbardlike electronic interactions among three-dimensional massless Dirac fermions that in a noninteracting system possess a global chiral U(1)$\otimes$SU(2) symmetry. We provide a lattice realization of such quasiparticle excitations in terms of the SLAC fermions, which should facilitate future numerical investigation of this subject using quantum Monte Carlo simulations~\cite{lauchli-SLAC, guo-maciejko-scaletter}, for example. We show that an isotropic interacting Dirac semimetal is described in terms of only \emph{four} linearly independent local quartic interactions. Beside studying the possible ordered or broken symmetry phases in this system, which set in through continuous quantum phase transitions, we also pay special attention to the restoration of partial and full chiral symmetry at various interacting fixed points. By performing a leading-order field theoretic RG analysis, controlled by a \emph{small} parameter $\epsilon=d-1$, about the lower-critical one spatial dimension ($d=1$), we find that an isotropic interacting chiral Dirac semimetal altogether supports \emph{nine} RG fixed points. One of them corresponds to the noninteracting trivial Gaussian fixed point, describing a stable Dirac semimetal for sufficiently weak, but generic short-range interactions. On the other hand, the system also supports four quantum critical (C$i$) and four bicritical (B$i$) fixed points at finite interaction couplings ($\sim \epsilon$), where $i=1,2,3,4$, see Table~\ref{tab:FPs}.

Even though we do not impose (either partial or full) chiral symmetry on the interacting theory $S_{\rm int}$ at the bare level, two QCPs, namely C1 and C4, transform as chiral U(1) scalars, while the remaining six fixed points (C2,C3), (B1,B2) and (B3,B4) pairwise transform as three two-component vectors under the chiral U(1) rotations. In addition, the pseudospin SU(2) chiral symmetry gets restored at three QCPs (C2, C3 and C4) and two BCPs (B3 and B4). Therefore, only one fixed point, namely C4, enjoys the full chiral U(1)$\otimes$SU(2) symmetry of the noninteracting systems, see Tables~\ref{tab:FPs}, ~\ref{tab:bilinears} and ~\ref{tab:AnomDim}.

The dynamic scaling exponent ($z$) and correlation length exponent ($\nu$) at all the QCPs are respectively $z=1$ and $\nu^{-1}=\epsilon$. Together they determine the scaling of the transition temperature $T_c \sim \delta^{\nu z}$ of the ordered states (up to a logarithmic correction due to the breakdown of the hyperscaling hypothesis in $d=3$), where $\delta$ is the reduced distance from a critical point. The value of $\nu=1/2$ is an exact result as the system resides at the upper critical three spatial dimension~\cite{zinn-justin:book, zinn-justin-moshe-moshe}. The momentum shell RG procedure although breaks the space-(imaginary)time Lorentz symmetry of the noninteracting system, it does not obscure the restoration of internal chiral symmetry at various RG fixed points.

We also demonstrate the imprints of some of these fixed points and emergent chiral symmetry among competing phases on a representative cut of the zero temperature global phase diagram, shown in Fig.~\ref{fig:phasediag}. This phase diagram displays an intriguing confluence of four competing mass orders, the scalar and pseudoscalar excitonic and superconducting masses, which are the energetically most favored ordered states at zero temperature as they uniformly and isotropically gap the Dirac point. In particular, we find high-symmetry lines in the phase diagram along which the chiral U(1) symmetry between two excitonic (scalar and pseudoscalar) masses and the pseudospin SU(2) symmetry among scalar or pseudoscalar excitonic and superconducting masses get restored. We also note that the phase diagram displays a chiral mirror symmetry about the 45$^\circ$ diagonal across which all the scalar and pseudoscalar mass orders transform into each other. Finally, it is worth pointing out that the arrangements among the competing and neighboring phases in the phase diagram are consistent with our previously proposed ``\emph{selection rule and organization principle}" in Ref.~\cite{szabo-moessner-roy}. In particular, one can immediately verify the following. (1) A quartic interaction $(\Psi^\dagger_{\rm Nam} \eta_\mu \Gamma_{\nu \rho} \Psi_{\rm Nam})^2$, written in the Nambu doubled basis $\Psi_{\rm Nam}$ [see Eq.~(\ref{eq:Nambudefinition})], is conducive for the nucleation of an ordered state, represented by the fermion bilinear $\Psi^\dag_{\rm Nam} {\rm O} \Psi_{\rm Nam}$, only if (a) ${\rm O} \equiv \eta_\mu \Gamma_{\nu \rho}$ or (b) $\{ {\rm O},\eta_\mu \Gamma_{\nu \rho} \}=0$. (2) Two ordered phases, represented by the fermion bilinears $\Psi^\dag_{\rm Nam} {\rm O}_1 \Psi_{\rm Nam}$ and $\Psi^\dag_{\rm Nam} {\rm O}_2 \Psi_{\rm Nam}$ reside next to each other only when $\{ {\rm O}_1, {\rm O}_2 \}=0$. The pseudospin SU(2) symmetry has also been discussed recently in the context of the $\eta$ pairing in Dirac and Weyl semimetals, as well as in nodal-loop semimetals~\cite{kaili-etapairing}. In the future we will demonstrate emergence of such symmetry from appropriate RG analysis in the context of extended Hubbard model in Weyl~\cite{roy-goswami-juricic:Weyl} and nodal-loop semimetals~\cite{broy-NLSM}.

Finally, we show that a weak anisotropy of the Dirac cone leaves the nature of various RG fixed points unchanged, except the fully U(1)$\otimes$SU(2) chiral symmetric critical point C4. Specifically, this critical point gets converted into a bicritical point even in a \emph{weakly} anisotropic Dirac system. A complete RG analysis, fixed point structure and the phase diagram in a three-dimensional anisotropic chiral Dirac semimetal is, however, left for a future investigation.

Here we address the role of electronic interactions and chiral symmetry restorations by performing a RG analysis about the lower-critical one spatial dimension. In the future, we will complement this analysis by performing an alternative RG analysis about the upper-critical three spatial dimension, with $\epsilon=3-d$, by accounting for order-parameter fluctuations within the framework of the Gross-Neveu-Yukawa theory~\cite{zinn-justin:book}. The existing methodology only allows to demonstrate the restoration of high-symmetry among the dominant mass orders~\cite{roy-goswami-juricic:mutlicriticality}. Therefore, a substantial generalization of the Gross-Neveu-Yukawa formalism is needed in order to demonstrate the restoration of partial or full chiral symmetry among all symmetry allowed fermion bilinears at RG fixed points, which therefore deserves a separate investigation. Schematically, all the fixed point in such a RG scheme are located at $\epsilon=0$, as the system resides exactly at the upper critical three spatial dimensions. However, before the coupling constants ultimately reach such Gaussian fixed points in the deep infrared regime, we expect the chiral symmetry (partial or full) to get restored (see Appendix~C of Ref.~\cite{roy-goswami-juricic:mutlicriticality}, for example).

\acknowledgments
B.R. was supported by the Startup grant from Lehigh University.

\pagebreak 

\appendix
%
%
\section{Symmetry classification of four-fermion interaction}~\label{seq:bilinears_class}

\begin{table}[t!]
\centering
\begin{tabular}{|c|c|c|c|c|}
\hline
bilinear & ${\mathcal P}$ & ${\mathcal T}$ & ${\mathcal C}$ & O(3)  \\
\hline
$\Psi^\dag \Gamma_{00} \Psi$ & $+$ & $+$ & $-$ & 0 \\
$\Psi^\dag \Gamma_{10} \Psi$ & $-$ & $+$ & $+$ & 0 \\
$\Psi^\dag \Gamma_{20} \Psi$ & $-$ & $-$ & $+$ & 0 \\
$\Psi^\dag \Gamma_{30} \Psi$ & $+$ & $+$ & $+$ & 0 \\
$\Psi^\dag \Gamma_{0j} \Psi$ & $+$ & $-$ & $+$ & 1 \\
$\Psi^\dag \Gamma_{1j} \Psi$ & $-$ & $-$ & $-$ & 1 \\
$\Psi^\dag \Gamma_{2j} \Psi$ & $-$ & $+$ & $-$ & 1 \\
$\Psi^\dag \Gamma_{3j} \Psi$ & $+$ & $-$ & $-$ & 1 \\
\hline
\end{tabular}
\caption{Classification of 16 fermion bilinears under the discrete parity ($\mathcal{P}$), time reversal ($\mathcal{T}$) and charge conjugation ($\mathcal{C}$) symmetries. Here $+$ ($-$) sign corresponds to even (odd) transformation of the bilinear, and $j=1,2,3$. The fifth column shows whether a fermion bilinear transforms as a scalar (0) or a three-component vector (1) under the spatial O(3) rotations, generated by $\{ \Gamma_{01}, \Gamma_{02}, \Gamma_{03} \}$.
}~\label{tab:class_bilinears}
\end{table}

The momentum-independent local four-fermion interactions are captured by the quartic terms of the form $(\Psi^\dag \Gamma_{\mu \nu} \Psi)(\Psi^\dag \Gamma_{\rho \lambda} \Psi)$, where $\mu,\nu,\rho,\lambda=0, \cdots, 3$. By imposing discrete parity ($\mathcal{P}$), time-reversal ($\mathcal{T}$), and charge conjugation ($\mathcal{C}$) symmetries, we reduce 136 possible interaction terms to the ones where both $\Psi^\dag \Gamma_{\mu \nu} \Psi$ and $\Psi^\dag \Gamma_{\rho \lambda} \Psi$ are either even or odd under $\mathcal{P}$, $\mathcal{T}$ and $\mathcal{C}$ separately, such that each quartic term is invariant under all three individual discrete symmetries. The classification of sixteen fermion bilinears under these three discrete symmetries is displayed in Table~\ref{tab:class_bilinears}. As there are no two identical rows in this table, there exists no interaction term that mixes any two different rows. Hence, the remaining four-fermion terms that are invariant under $\mathcal{P}$, $\mathcal{T}$ and $\mathcal{C}$ are
\begin{align}
&(\Psi^\dag \Gamma_{\mu \nu} \Psi)^2 &  &\mathrm{(16\ of\ them),}\label{eq:square_terms}\\
&(\Psi^\dag \Gamma_{\mu j} \Psi)(\Psi^\dag \Gamma_{\mu k} \Psi) &  &\mathrm{(12\ of\ them)}, \label{eq:mixed_terms}
\end{align}
where $j \neq k= 1,2,3$.

The number of quartic terms is further reduced when we invoke the spatial rotational symmetry. First of all, all terms from (\ref{eq:mixed_terms}) get eliminated, as they do not transform as scalars nor as vectors under the O(3) spatial rotations. Furthermore, it organizes (\ref{eq:square_terms}) into four scalars (O(3) vectors) for $\Gamma_{\mu 0}$ ($\Gamma_{\mu j}$), with $j=1,2,3$. Therefore, the eight symmetry allowed four-fermion terms are of the form
\begin{equation}
(\Psi^\dag \Gamma_{\mu 0} \Psi)^2{\rm \ and\ }\sum_{j=1}^3(\Psi^\dag \Gamma_{\mu j} \Psi)^2.
\end{equation}
In Eq.~(\ref{eq:Lint}) the matrices $\Gamma_{\mu 0}$ and $\Gamma_{\mu j}$ appear with couplings $g^s_{_\mu}$, $g^t_{_\mu}$ respectively. We note that each component of $\Gamma_{\mu 0}$ ($\Gamma_{\mu j}$) for $\mu=0, \cdots, 3$ transforms as scalar (three-component vector) under the spatial O(3) rotations, generated by $\{\Gamma_{01}, \Gamma_{02}, \Gamma_{03} \}$. As the O(3) group is \emph{isomorphic} to SU(2), we can immediately conclude that the total number of linearly independent quartic terms in an isotropic Dirac semimetal is \emph{four}, the dimensionality of the vectors $\Gamma_{\mu 0}$ and $\Gamma_{\mu j}$. Furthermore, we can choose four O(3) scalar quartic terms $(\Psi^\dag \Gamma_{\mu 0} \Psi)^2$ as the linearly independent four-fermion interactions. Next we explicitly demonstrate these outcomes using the Fierz identity.

\section{Fierz reduction of four-fermion interaction}~\label{sec:Fierz}

Given a complete basis of Hermitian matrices, the Fierz identity allows us to express any four-fermion term $(\Psi^\dag \Gamma_{\mu \nu} \Psi)(\Psi^\dag \Gamma_{\rho \lambda} \Psi)$ as a linear combination of the others. Schematically, the Fierz relation can be written as 
\begin{align}~\label{eq:Fierz}
(\Psi^\dag \Gamma_{\mu \nu} \Psi)&(\Psi^\dag \Gamma_{\rho \lambda} \Psi)=   
-\frac{1}{16}\sum_{\alpha,\beta,\gamma,\delta} \Tr(\Gamma_{\mu \nu} \Gamma_{\alpha \beta} \Gamma_{\rho \lambda} \Gamma_{\gamma \delta})
\times (\Psi^\dag \Gamma_{\alpha \beta} \Psi)(\Psi^\dag \Gamma_{\gamma \delta} \Psi),
\end{align}
where $\alpha,\beta,\gamma,\delta=0,\cdots, 3$. Next we demonstrate the general principle to find linearly independent quartic terms in a given interacting model. For example, if $X$ is an array of the four-fermion terms, then the above Fierz relation can be cast as 
\begin{equation}
X=MX \Rightarrow (M-{\bm 1})X \Rightarrow FX=0,
\end{equation}
where $M$ contains the linear connections among the quartic terms due to Eq.~(\ref{eq:Fierz}), and $F=M-{\bm 1}$ is the Fierz matrix. The number of linearly independent four-fermion terms is equal to ${\rm D}(F)-{\rm R}(F)$, where ${\rm D}(F)$ and ${\rm R}(F)$ are respectively the dimension and rank of the square matrix $F$. Since the isotropic and anisotropic DSM are bestowed with different symmetries, the explicit forms of the corresponding Fierz matrices and the numbers of independent quartic terms in these two cases are distinct. Therefore, we present the two cases separately following the general approach outlined above.

\subsection{Isotropic Dirac semimetal}

For the isotropic Dirac semimetal there are eight symmetry allowed quartic terms, see Eq.~(\ref{eq:Lint}), which can be organized into $X$ according to 
\begin{align}
X^\top = \Big[&(\Psi^\dag \Gamma_{00} \Psi)^2,
(\Psi^\dag \Gamma_{10} \Psi)^2,
(\Psi^\dag \Gamma_{20} \Psi)^2,
(\Psi^\dag \Gamma_{30} \Psi)^2, \nonumber \\
&\sum_{j=1}^3(\Psi^\dag \Gamma_{0j} \Psi)^2,
\sum_{j=1}^3(\Psi^\dag \Gamma_{1j} \Psi)^2,
\sum_{j=1}^3(\Psi^\dag \Gamma_{2j} \Psi)^2,
\sum_{j=1}^3(\Psi^\dag \Gamma_{3j} \Psi)^2 \Big]. 
\end{align}
The corresponding eight-dimensional Fierz matrix reads as
\begin{align}
F_{\rm iso}=
\left(
\begin{array}{cccccccc}
 5 & 1 & 1 & 1 & 1 & 1 & 1 & 1 \\
 1 & 5 & -1 & -1 & 1 & 1 & -1 & -1 \\
 1 & -1 & 5 & -1 & 1 & -1 & 1 & -1 \\
 1 & -1 & -1 & 5 & 1 & -1 & -1 & 1 \\
 3 & 3 & 3 & 3 & 3 & -1 & -1 & -1 \\
 3 & 3 & -3 & -3 & -1 & 3 & 1 & 1 \\
 3 & -3 & 3 & -3 & -1 & 1 & 3 & 1 \\
 3 & -3 & -3 & 3 & -1 & 1 & 1 & 3 \\
\end{array}
\right),
\end{align}
with ${\rm R}(F_{\rm iso})=4$. Therefore, the number of independent coupling constants is $\mathrm{D}(F_{\rm iso})-\mathrm{R}(F_{\rm iso})=4$. Without any loss of generality, we choose the four single-component quartic terms containing $\Gamma_{\mu 0}$, each of which transforms as a scalar under spatial O(3) rotations, as the independent ones, see Eq.~(\ref{eq:S_int}). The remaining four quartic terms containing $\Gamma_{\mu j}$, which for any given $\mu$ transform as a three-component vector under spatial O(3) rotations, can be expressed as 
\allowdisplaybreaks[4]
\begin{align}
\sum_{j=1}^3(\Psi^\dag \Gamma_{0j} \Psi)^2 =
-2 (\Psi^\dag \Gamma_{00} \Psi)^2 - (\Psi^\dag \Gamma_{10} \Psi)^2 - (\Psi^\dag \Gamma_{20} \Psi)^2 - (\Psi^\dag \Gamma_{30} \Psi)^2, \nonumber \\
\sum_{j=1}^3(\Psi^\dag \Gamma_{1j} \Psi)^2 =
-(\Psi^\dag \Gamma_{00} \Psi)^2 -2 (\Psi^\dag \Gamma_{10} \Psi)^2 + (\Psi^\dag \Gamma_{20} \Psi)^2 + (\Psi^\dag \Gamma_{30} \Psi)^2, \nonumber \\
\sum_{j=1}^3(\Psi^\dag \Gamma_{2j} \Psi)^2 =
-(\Psi^\dag \Gamma_{00} \Psi)^2 + (\Psi^\dag \Gamma_{10} \Psi)^2 -2 (\Psi^\dag \Gamma_{20} \Psi)^2 + (\Psi^\dag \Gamma_{30} \Psi)^2, \nonumber\\
\sum_{j=1}^3(\Psi^\dag \Gamma_{3j} \Psi)^2 =
-(\Psi^\dag \Gamma_{00} \Psi)^2 + (\Psi^\dag \Gamma_{10} \Psi)^2 + (\Psi^\dag \Gamma_{20} \Psi)^2 -2 (\Psi^\dag \Gamma_{30} \Psi)^2.
\end{align}
Therefore, whenever we generate any one of these quartic terms through the quantum loop corrections, it is expressed in terms of the four quartic terms appearing in Eq.~(\ref{eq:S_int}). Furthermore, the fact that an interacting isotropic chiral Dirac semimetal can be described by only four linearly independent quartic terms is consistent with the existence of four independent superconducting orders, tabulated in Table~\ref{tab:bilinears} (last four rows).

\subsection{Anisotropic Dirac semimetals}

When we introduce an anisotropy between the in-plane ($v_x$ and $v_y$, with $v_x=v_y=v_\perp$) and the perpendicular or out of the plane ($v_z$) components of the Fermi velocity (germane in a tetragonal system), each three-component quartic term splits into two, with the corresponding coupling constants splitting as $g^t_{_\mu} \to(g^\perp_{_\mu},g^z_{_\mu})$ for $\mu=0, \cdots, 3$, yielding the interacting Lagrangian
\begin{align}
L^{\rm ani}_{\mathrm{int}}=
&g^s_{_0}(\Psi^\dag \Gamma_{00} \Psi)^2 +
g^s_{_1}(\Psi^\dag \Gamma_{10} \Psi)^2 + g^s_{_2}(\Psi^\dag \Gamma_{20} \Psi)^2 + 
g^s_{_3}(\Psi^\dag \Gamma_{30} \Psi)^2 \nonumber  \\
+&g_{_0}^\perp\sum_{j=1}^2(\Psi^\dag \Gamma_{0j} \Psi)^2+ g_{_0}^z(\Psi^\dag \Gamma_{03} \Psi)^2+
g_{_1}^\perp\sum_{j=1}^2(\Psi^\dag \Gamma_{1j} \Psi)^2+ g_{_1}^z(\Psi^\dag \Gamma_{13} \Psi)^2 \nonumber  \\
+&g_{_2}^\perp\sum_{j=1}^2(\Psi^\dag \Gamma_{2j} \Psi)^2+ g_{_2}^z(\Psi^\dag \Gamma_{23} \Psi)^2+  
+g_{_3}^\perp\sum_{j=1}^2(\Psi^\dag \Gamma_{3j} \Psi)^2 + g_{_3}^z(\Psi^\dag \Gamma_{33} \Psi)^2. 
\end{align} 
Hence the array ($X$) containing all the quartic terms reads
\begin{align}
X^\top = \Bigg[&
(\Psi^\dag \Gamma_{00} \Psi)^2,
(\Psi^\dag \Gamma_{10} \Psi)^2,
(\Psi^\dag \Gamma_{20} \Psi)^2,
(\Psi^\dag \Gamma_{30} \Psi)^2,
\sum_{j=1}^2(\Psi^\dag \Gamma_{0j} \Psi)^2,
(\Psi^\dag \Gamma_{03} \Psi)^2, \nonumber \\
&\sum_{j=1}^2(\Psi^\dag \Gamma_{1j} \Psi)^2,(\Psi^\dag \Gamma_{13} \Psi)^2,
\sum_{j=1}^2(\Psi^\dag \Gamma_{2j} \Psi)^2,(\Psi^\dag \Gamma_{23} \Psi)^2,
\sum_{j=1}^2(\Psi^\dag \Gamma_{3j} \Psi)^2, (\Psi^\dag \Gamma_{33} \Psi)^2 \Bigg].
\end{align}
The twelve-dimensional Fierz matrix then reads
\begin{align}
F_{\mathrm{ani}}=
\left(
\begin{array}{cccccccccccc}
 5 & 1 & 1 & 1 & 1 & 1 & 1 & 1 & 1 & 1 & 1 & 1 \\
 1 & 5 & -1 & -1 & 1 & 1 & 1 & 1 & -1 & -1 & -1 & -1 \\
 1 & -1 & 5 & -1 & 1 & 1 & -1 & -1 & 1 & 1 & -1 & -1 \\
 1 & -1 & -1 & 5 & 1 & 1 & -1 & -1 & -1 & -1 & 1 & 1 \\
 2 & 2 & 2 & 2 & 4 & -2 & 0 & -2 & 0 & -2 & 0 & -2 \\
 1 & 1 & 1 & 1 & -1 & 5 & -1 & 1 & -1 & 1 & -1 & 1 \\
 2 & 2 & -2 & -2 & 0 & -2 & 4 & -2 & 0 & 2 & 0 & 2 \\
 1 & 1 & -1 & -1 & -1 & 1 & -1 & 5 & 1 & -1 & 1 & -1 \\
 2 & -2 & 2 & -2 & 0 & -2 & 0 & 2 & 4 & -2 & 0 & 2 \\
 1 & -1 & 1 & -1 & -1 & 1 & 1 & -1 & -1 & 5 & 1 & -1 \\
 2 & -2 & -2 & 2 & 0 & -2 & 0 & 2 & 0 & 2 & 4 & -2 \\
 1 & -1 & -1 & 1 & -1 & 1 & 1 & -1 & 1 & -1 & -1 & 5 \\
\end{array}
\right),
\end{align}
and now $\mathrm{R}(F_{\mathrm{ani}})=7$. Therefore, we have five linearly independent quartic terms. The additional coupling constant (besides $g^s_{_\mu}$ with $\mu=0, \cdots, 3$) can be chosen to be $g_{_1}^z$, for example, see Sec.~\ref{subsec:anisotropy}. The remaining seven quartic terms are then given by
\allowdisplaybreaks[4]
\begin{align}
\sum_{j=1}^2(\Psi^\dag \Gamma_{0j} \Psi)^2 =& -(\Psi^\dag \Gamma_{00} \Psi)^2 -(\Psi^\dag \Gamma_{20} \Psi)^2 - (\Psi^\dag \Gamma_{30} \Psi)^2
+(\Psi^\dag \Gamma_{13} \Psi)^2, \nonumber\\
(\Psi^\dag \Gamma_{03} \Psi)^2 =& - (\Psi^\dag \Gamma_{00} \Psi)^2 - (\Psi^\dag \Gamma_{10} \Psi)^2 - (\Psi^\dag \Gamma_{13} \Psi)^2, \nonumber\\
\sum_{j=1}^2(\Psi^\dag \Gamma_{1j} \Psi)^2 =&
-(\Psi^\dag \Gamma_{00} \Psi)^2 -2(\Psi^\dag \Gamma_{10} \Psi)^2 + (\Psi^\dag \Gamma_{20} \Psi)^2 + (\Psi^\dag \Gamma_{30} \Psi)^2
-(\Psi^\dag \Gamma_{13} \Psi)^2, \nonumber \\
\sum_{j=1}^2(\Psi^\dag \Gamma_{2j} \Psi)^2 =& - (\Psi^\dag \Gamma_{00} \Psi)^2 - (\Psi^\dag \Gamma_{20} \Psi)^2 + (\Psi^\dag \Gamma_{30} \Psi)^2 - (\Psi^\dag \Gamma_{13} \Psi)^2, \nonumber\\
(\Psi^\dag \Gamma_{23} \Psi)^2 =&
(\Psi^\dag \Gamma_{10} \Psi)^2 - (\Psi^\dag \Gamma_{20} \Psi)^2
+(\Psi^\dag \Gamma_{13} \Psi)^2, \nonumber\\
\sum_{j=1}^2(\Psi^\dag \Gamma_{3j} \Psi)^2 =&
-(\Psi^\dag \Gamma_{00} \Psi)^2 + (\Psi^\dag \Gamma_{20} \Psi)^2 - (\Psi^\dag \Gamma_{30} \Psi)^2 - (\Psi^\dag \Gamma_{13} \Psi)^2, \nonumber\\
(\Psi^\dag \Gamma_{33} \Psi)^2 =& (\Psi^\dag \Gamma_{10} \Psi)^2 - (\Psi^\dag \Gamma_{30} \Psi)^2 + (\Psi^\dag \Gamma_{13} \Psi)^2. 
\end{align}
The fact that an interacting anisotropic (tetragonal) Dirac semimetal is described by five linearly independent quartic terms can be justified from the fact that a Dirac system with reduced in-plane rotational symmetry supports five independent local pairings, as the three-component vector pairing $\Delta_2^p$ (see Table~\ref{tab:bilinears}) splits into the in-plane (with $j=1,2$) and out of plane (with $j=3$) components. Similarly, one can show that in orthorombic system ($v_x \neq v_y \neq v_z$), an interacting Dirac semimetal is described by six linearly independent local four-fermion interactions.

{\bf Open Access}. This article is distributed under the terms of the Creative Commons Attribution License (CC-BY 4.0), which permits any use, distribution and reproduction in any medium, provided the original author(s) and source are credited.




\begin{thebibliography}{}
 
\bibitem{peskin-schroeder} M. E. Peskin and D. V. Schroeder, {\it An Introduction to Quantum Field Theory} (Addison-Wesley, Reading, MA, 1995).

\bibitem{Shen-book}  S. Q. Shen, \emph{Topological Insulators-Dirac Equation in Condensed Matters} (Springer, New York, 2012).

\bibitem{Bernevig-book} Bernevig, B. A., and T. L. Hughes, \emph{Topological insulators and topological superconductors} (Princeton University Press, Princeton, NJ, 2013).

\bibitem{balatsky:review} T. O. Wehling, A. M. Black-Schaffer, A. V. Balatsky, \emph{Dirac Materials}, Adv. Phys. {\bf 63}, 1 (2014).

\bibitem{armitage:RMP} N. P. Armitage, E. J. Mele, and A. Vishwanath, \emph{Weyl and Dirac semimetals in three-dimensional solids}, Rev. Mod. Phys. {\bf 90}, 015001 (2018).

\bibitem{graphene:RMP} A. H. Castro Neto, F. Guinea, N. M. R. Peres, K. S. Novoselov, and A. K. Geim, \emph{The electronic properties of graphene}, Rev. Mod. Phys. {\bf 81}, 109 (2009).

\bibitem{cd2as3:Exp} S. Borisenko, Q. Gibson, D. Evtushinsky, V. Zabolotnyy, B.Buchner, and R. J. Cava, \emph{Experimental Realization of a Three-Dimensional Dirac Semimetal}, Phys. Rev. Lett. {\bf 113}, 027603(2014).

\bibitem{na3bi:Exp} Z. K. Liu, B. Zhou, Z. J. Wang, H. M. Weng, D. Prabhakaran,S.-K. Mo, Y. Zhang, Z. X. Shen, Z. Fang, X. Dai, Z. Hussain,and Y. L. Chen, \emph{Discovery of a Three-Dimensional Topological Dirac Semimetal}, Na$_{\rm 3}$Bi, Science {\bf 343}, 864(2014).

\bibitem{anderson} P. W. Anderson, \emph{More Is Different}, Science {\bf 177}, 393 (1972).

\bibitem{zinn-justin:book} J. Zinn-Justin, \emph{Quantum Field Theory and Critical Phenomena} (Oxford Science, Oxford, 2002).

\bibitem{zinn-justin-moshe-moshe} M. Moshe and J. Zinn-Justin, \emph{Quantum field theory in the large N limit: a review}, Phys. Rep. {\bf 385}, 69 (2003).

\bibitem{peccei-quinn} R. D. Peccei and H. R. Quinn, \emph{{\bf CP} Conservation in the Presence of Pseudoparticles}, Phys. Rev. Lett. {\bf 38}, 1440 (1977).

\bibitem{weinberg} S. Weinberg, \emph{A New Light Boson?}, Phys. Rev. Lett. {\bf 40}, 223(1978).

\bibitem{wilczek} F. Wilczek, \emph{Problem of Strong {\bf P} and {\bf T} Invariance in the Presence of Instantons}, Phys. Rev. Lett. {\bf 40}, 279(1978).

\bibitem{roy-foster:PRX} B. Roy and M. S. Foster, \emph{Quantum Multicriticality near the Dirac-Semimetal to Band-Insulator Critical Point in Two Dimensions: A Controlled Ascent from One Dimension}, Phys. Rev. X {\bf 8}, 011049 (2018).

\bibitem{Kirkpatrick-Belitz:dopedDSM} T. R. Kirkpatrick and D. Belitz, \emph{Soft modes and nonanalyticities in a clean Dirac metal}, Phys. Rev. B {\bf 99} 085109 (2019).

\bibitem{nielsen} H. B. Nielsen and M. Ninomiya, \emph{A no-go theorem for regularizing chiral fermions}, Phys. Lett. B {\bf 105}, 219(1981).

\bibitem{lauchli-SLAC} T. C. Lang and A. M.  L\"auchli, \emph{Quantum Monte Carlo Simulation of the Chiral Heisenberg Gross-Neveu-Yukawa Phase Transition with a Single Dirac Cone}, Phys. Rev. Lett. {\bf 123}, 137602 (2019).

\bibitem{guo-maciejko-scaletter} Y. Huang, H. Guo, J. Maciejko, R. T. Scalettar, and S. Feng, \emph{Antiferromagnetic transitions of Dirac fermions in three dimensions}, Phys. Rev. B {\bf 102}, 155152 (2020).

\bibitem{quinn-weinstein} H. R. Quinn and M. Weinstein, \emph{Lattice theories of chiral fermions}, Phys. Rev. D {\bf 34}, 2440 (1986).

\bibitem{wilson} K. G. Wilson, \emph{Confinement of quarks}, Phys. Rev. D {\bf 10}, 2445 (1974).

\bibitem{kogut-susskind} J. Kogut and L. Susskind, \emph{Hamiltonian formulation of Wilson's lattice gauge theories}, Phys. Rev. D {\bf 11}, 395 (1975).

\bibitem{roy-goswami-sau} B. Roy, P. Goswami and J. D. Sau, \emph{Continuous and discontinuous topological quantum phase transitions},
Phys. Rev. B {\bf 94}, 041101(R) (2016).

\bibitem{nason} P. Nason, \emph{The lattice Schwinger model with SLAC fermions}, Nucl. Phys. B {\bf 260}, 269 (1985).

\bibitem{costella} J. P. Costella, \emph{A new proposal for the fermion doubling problem}, arXiv:hep-lat/0207008




\bibitem{goswami-chakravarty} P. Goswami and S. Chakravarty, \emph{Quantum Criticality between Topological and Band Insulators in 3+1 Dimensions}, Phys. Rev. Lett. {\bf 107}, 196803(2011).

\bibitem{isobe-nagaosa} H. Isobe and N. Nagaosa, \emph{Theory of a quantum critical phenomenon in a topological insulator: (3+1)-dimensional quantum electrodynamics in solids}, Phys. Rev. B {\bf 86}, 165127 (2012).

\bibitem{jose-gonzalez} J. Gonz\'alez, \emph{Phase diagram of the quantum electrodynamics of two-dimensional and three-dimensional Dirac semimetals}, Phys. Rev. B {\bf 92}, 125115 (2015).

\bibitem{prokofev} I. S. Tupitsyn and N. V. Prokof'ev, \emph{Stability of Dirac Liquids with Strong Coulomb Interaction}, Phys. Rev. Lett. {\bf 118}, 026403 (2017).

\bibitem{dassarma-throckmorton} R. E. Throckmorton, J. Hofmann, E. Barnes, and S. Das Sarma, \emph{Many-body effects and ultraviolet renormalization in three-dimensional Dirac materials}, Phys. Rev. B {\bf 92}, 115101 (2015).

\bibitem{juricic-herbut-semenoff} V. Juri\u ci\' c, I. F. Herbut, and G. W. Semenoff, \emph{Coulomb interaction at the metal-insulator critical point in graphene}, Phys. Rev. B {\bf 80}, 081405(R) (2009). 

\bibitem{drut-lahde} J. E. Drut and T. A. L\"ahde, \emph{Is Graphene in Vacuum an Insulator?}, Phys. Rev. Lett. {\bf 102}, 026802 (2009).

\bibitem{roy-dassarma} B. Roy and S. Das Sarma, \emph{Quantum phases of interacting electrons in three-dimensional dirty Dirac semimetals},
Phys. Rev. B {\bf 94}, 115137 (2016).

\bibitem{katslenson} V. V. Braguta, M. I. Katsnelson, A. Yu. Kotov, and A. A. Nikolaev, \emph{Monte Carlo study of Dirac semimetals phase diagram}, Phys. Rev. B {\bf 94}, 205147 (2016).

\bibitem{zhao-wang:Coulomb} P-L. Zhao and A-M. Wang, \emph{Interplay between tilt, disorder, and Coulomb interaction in type-I Dirac fermions}, Phys. Rev. B {\bf 100} 125138 (2019).

\bibitem{yang-wang-liu:Coulomb} Z.-K. Yang, J.-R. Wang, and G-Z. Liu, \emph{Effects of Dirac cone tilt in a two-dimensional Dirac semimetal}, Phys. Rev. B {\bf 98} 195123 (2018).



 

\bibitem{maciejko-nandkishore:Weyl} J. Maciejko and R. Nandkishore, \emph{Weyl semimetals with short-range interactions}, Phys. Rev. B {\bf 90}, 035126(2014).

\bibitem{roy-goswami-juricic:Weyl} B. Roy, P. Goswami, and V. Juri\u ci\' c, \emph{Interacting Weyl fermions: Phases, phase transitions, and global phase diagram}, Phys. Rev. B {\bf 95}, 201102(R) (2017). 

\bibitem{herbut-juricic-roy} I. F. Herbut, V. Juri\u ci\' c and B. Roy, \emph{Theory of interacting electrons on the honeycomb lattice}, Phys. Rev. B {\bf 79}, 085116 (2009).

\bibitem{szabo-moessner-roy} A. L. Szab\' o, R. Moessner, and B. Roy, \emph{Interacting spin-3/2 fermions in a Luttinger (semi)metal: competing phases and their selection in the global phase diagram}, arXiv:1811.12415

\bibitem{roy-ghorashi-foster-nevidomskyy} B. Roy, S. A. A. Ghorashi, M. S. Foster, and A. H. Nevidomskyy, \emph{Topological superconductivity of spin-3/2 carriers in a three-dimensional doped Luttinger semimetal}, Phys. Rev. B {\bf 99}, 054505 (2019).

\bibitem{pseudospin-1} C. N. Yang and S. C. Zhang, \emph{SO$_{\rm 4}$ symmetry in a Hubbard model}, Mod. Phys. Lett. B {\bf 04}, 759 (1990).

\bibitem{pseudospin-2} A. Auerbach, \emph{Interacting Electrons and Quantum Magnetism} (Springer-Verlag, New York, NY, 1994).

\bibitem{pseudospin-3} M. Hermele, \emph{SU(2) gauge theory of the Hubbard model and application to the honeycomb lattice}, Phys. Rev. B {\bf 76}, 035125(2007).

\bibitem{ohsaku} T. Ohsaku, \emph{BCS and generalized BCS superconductivity in relativistic quantum field theory: Formulation}, Phys. Rev. B {\bf 65}, 024512 (2001).

\bibitem{fu-berg} L. Fu and E. Berg, \emph{Odd-Parity Topological Superconductors: Theory and Application to ${\rm Cu_x Bi_2 Se_3}$}, Phys. Rev. Lett. {\bf 105}, 097001 (2010).

\bibitem{kaili-etapairing} K. Li, \emph{$\eta$-pairing in correlated fermion models with spin-orbit coupling}, Phys. Rev. B {\bf 102}, 165150 (2020). 

\bibitem{broy-NLSM} B. Roy, \emph{Interacting nodal-line semimetal: Proximity effect and spontaneous symmetry breaking}, Phys. Rev. B {\bf 96}, 041113(R) (2017).   

\bibitem{roy-goswami-juricic:mutlicriticality} B. Roy, P. Goswami, and V. Juri\u ci\' c, \emph{Itinerant quantum multicriticality of two-dimensional Dirac fermions}, Phys. Rev. B {\bf 97}, 205117 (2018).   
 
 
 

\end{thebibliography}
\end{document}